\documentclass{article}%
\usepackage{graphicx}
\usepackage{amsmath}
\usepackage{amsfonts}
\usepackage{amssymb}%
\newtheorem{theorem}{Theorem}
{}

\newtheorem{lemma}{Lemma}
{}

\newtheorem{remark}{Remark}

\newenvironment{proof}[1][Proof]{\textbf{#1.} }{\ \rule{0.5em}{0.5em}}

\begin{document}

\author{O. A. Veliev\\{\small \ Dept. of Math, Fen-Ed. Fak, Dogus University.,}\\{\small Acibadem, Kadikoy, Istanbul, Turkey,}\\{\small \ e-mail: oveliev@dogus.edu.tr}}
\title{\textbf{Asymptotic Analysis of the Periodic Schrodinger Operator}}
\date{}
\maketitle

\begin{abstract}
In this paper we obtain asymptotic formulas of arbitrary order for the Bloch
eigenvalues and Bloch functions of the Schrodinger operator $-\Delta+q(x),$ of
arbitrary dimension, with periodic, with respect to arbitrary lattice,
potential $q(x)$. Moreover, we estimate the measure of the isoenergetic
surfaces in the high energy region.

\end{abstract}

\section{Introduction}

\bigskip In this paper we consider the operator%
\begin{equation}
L(q(x))=-\Delta+q(x),\ x\in\mathbb{R}^{d},\ d\geq2
\end{equation}
with a periodic (relative to a lattice $\Omega$) potential $q(x)\in W_{2}%
^{s}(F),$ where

$s\geq s_{0}=\frac{3d-1}{2}(3^{d}+d+2)+\frac{1}{4}d3^{d}+d+6,$ $F\equiv
\mathbb{R}^{d}/\Omega$ is a fundamental domain of $\Omega.$ Without loss of
generality it can be assumed that the measure $\mu(F)$ of $F$ is $1$ and
$\int_{F}q(x)dx=0.$ Let $L_{t}(q(x))$ be the operator generated in $F$ by (1)
and the conditions:%
\begin{equation}
u(x+\omega)=e^{i(t,\omega)}u(x),\ \forall\omega\in\Omega,
\end{equation}
where $t\in F^{\star}\equiv\mathbb{R}^{d}/\Gamma$ and $\Gamma$ is the lattice
dual to $\Omega$, that is, $\Gamma$ is the set of all vectors $\gamma
\in\mathbb{R}^{d}$ satisfying $(\gamma,\omega)\in2\pi Z$ for all $\omega
\in\Omega.$ It is well-known that ( see [2]) the spectrum of the operator
$L_{t}(q(x))$ consists of the eigenvalues

$\Lambda_{1}(t)\leq\Lambda_{2}(t)\leq....$The function $\Lambda_{n}(t)$ is
called $n$-th band function and its range $A_{n}=\left\{  \Lambda_{n}(t):t\in
F^{\ast}\right\}  $ is called the $n$-th band of the spectrum $Spec(L)$ of $L$
and $Spec(L)=\cup_{n=1}^{\infty}A_{n}$. The eigenfunction $\Psi_{n,t}(x)$ of
$L_{t}(q(x))$ corresponding to the eigenvalue $\Lambda_{n}(t)$ is known as
Bloch functions. In the case $q(x)=0$ these eigenvalues and eigenfunctions are
$\mid\gamma+t\mid^{2}$ and $e^{i(\gamma+t,x)}$ for $\gamma\in\Gamma$.

This paper consists of 4 section. First section is the introduction, where we
describe briefly the scheme of this paper and discuss the related papers.

In papers [13-17] for the first time the eigenvalues $\left\vert
\gamma+t\right\vert ^{2}$, for big $\ \gamma\in\Gamma,$ were divided into two
groups: non-resonance ones and resonance ones and for the perturbations of
each group various asymptotic formulae were obtained. Let the potential $q(x)$
be a trigonometric polynomial%

\[
\sum_{\gamma\in Q}q_{\gamma}e^{i(\gamma,x)},
\]
where $q_{\gamma}=(q(x),e^{i(\gamma,x)})=\int_{F}q(x)e^{-i(\gamma_{1},x)}dx,$
and $Q$ consists of a finite number of vectors $\gamma$ from $\Gamma.$ Then
the eigenvalue $\left\vert \gamma+t\right\vert ^{2}$ \ is called a
non-resonance eigenvalue if $\gamma+t$ does not belong to any of the sets

$\{x\in\mathbb{R}^{d}:\mid\mid x\mid^{2}-\mid x+b\mid^{2}\mid<\mid
x\mid^{\alpha_{1}}\},$ that is, if $\gamma+t$ lies far from the diffraction
hyperplanes $\{x\in\mathbb{R}^{d}:\mid x\mid^{2}=\mid x+b\mid^{2}\},$ where
$\alpha_{1}\in(0,1),$

$b\in\{b_{1}+b_{2}+...b_{m}:b_{1},b_{2},...b_{m}\in Q\},$ and $m$ is fixed
integer (see [15-17]).

If $q(x)\in W_{2}^{s}(F),$ then to describe the non-resonance and resonance
eigenvalues $\left\vert \gamma+t\right\vert ^{2}$ of the order of $\rho^{2}$ (
written as $\left\vert \gamma+t\right\vert ^{2}\sim\rho^{2}$) for big
paramater $\rho$ we write the potential $q(x)\in W_{2}^{s}(F)$ in the form
\begin{equation}
q(x)=\sum_{\gamma_{1}\in\Gamma(\rho^{\alpha})}q_{\gamma_{1}}e^{i(\gamma
_{1},x)}+O(\rho^{-p\alpha}),
\end{equation}
where $\Gamma(\rho^{\alpha})=\{\gamma\in\Gamma:0<$ $\mid\gamma\mid
<\rho^{\alpha})\}$, $p=s-d,$ $\alpha=\frac{1}{q},$ $q=3^{d}+d+2,$ and the
relation $\left\vert \gamma+t\right\vert ^{2}\sim\rho^{2}$ means that
$c_{1}\rho<\left\vert \gamma+t\right\vert <c_{2}\rho$. Here and in subsequent
relations we denote by $c_{i}$ ($i=1,2,...)$ the positive, independent of
$\rho$ constants whose exact values are inessential. Note that $q(x)\in
W_{2}^{s}(F)$ means that $\sum_{\gamma}\mid q_{\gamma}\mid^{2}(1+\mid
\gamma\mid^{2s})<\infty.$ If $s\geq d,$ then
\begin{equation}
\sum_{\gamma}\mid q_{\gamma}\mid<c_{3},\text{ }\sup\mid\sum_{\gamma
\notin\Gamma(\rho^{\alpha})}q_{\gamma}e^{i(\gamma,x)}\mid\leq\sum_{\mid
\gamma\mid\geq\rho^{\alpha}}\mid q_{\gamma}\mid=O(\rho^{-p\alpha}),
\end{equation}
i.e., (3) holds. By definition, put $\alpha_{k}=3^{k}\alpha$ for $k=1,2,...$
and introduce the sets $V_{\gamma_{1}}(\rho^{\alpha_{1}})\equiv\{x\in
\mathbb{R}^{d}:\mid\mid x\mid^{2}-\mid x+\gamma_{1}\mid^{2}\mid<\rho
^{\alpha_{1}}\},$%
\[
E_{1}(\rho^{\alpha_{1}},p)\equiv\bigcup_{\gamma_{1}\in\Gamma(p\rho^{\alpha}%
)}V_{\gamma_{1}}(\rho^{\alpha_{1}}),\text{ }U(\rho^{\alpha_{1}},p)\equiv
\mathbb{R}^{d}\backslash E_{1}(\rho^{\alpha_{1}},p)
\]%
\[
E_{k}(\rho^{\alpha_{k}},p)\equiv\bigcup_{\gamma_{1},\gamma_{2},...,\gamma
_{k}\in\Gamma(p\rho^{\alpha})}(\cap_{i=1}^{k}V_{\gamma_{i}}(\rho^{\alpha_{k}%
})),
\]
where the intersection $\cap_{i=1}^{k}V_{\gamma_{i}}$ in the definition of
$E_{k}$ is taken over $\gamma_{1},\gamma_{2},...,\gamma_{k},$ that are
linearly independent. The set $U(\rho^{\alpha_{1}},p)$ is said to be a
non-resonance domain and the eigenvalue $\left\vert \gamma+t\right\vert
^{2}\sim\rho^{2}$ is called a non-resonance eigenvalue if $\gamma+t\in
U(\rho^{\alpha_{1}},p).$ The domains $V_{\gamma_{1}}(\rho^{\alpha_{1}})$ for
$\gamma_{1}\in\Gamma(p\rho^{\alpha})$ are called resonance domains and
$\mid\gamma+t\mid^{2}$ is called a resonance eigenvalue if $\gamma+t\in
V_{\gamma_{1}}(\rho^{\alpha_{1}}).$

In section 2 we prove that for each $\gamma+t\in U(\rho^{\alpha_{1}},p)$ there
exists an eigenvalue $\Lambda_{N}(t)$ of the operator $L_{t}(q(x))$ satisfying
the following formulae
\begin{equation}
\Lambda_{N}(t)=\mid\gamma+t\mid^{2}+F_{k-1}(\gamma+t)+O(\mid\gamma
+t\mid^{-3k\alpha})
\end{equation}
for $k=1,2,...,[\frac{1}{3}(p-\frac{1}{2}q(d-1))],$ where $[a]$ denotes the
integer part of $a,$ $F_{0}=0,$ and $F_{k-1}$ ( for $k>1)$ is explicitly
expresed by the potential $q(x)$ and eigenvalues of $L_{t}(0).$ Besides, we
prove that if the conditions
\begin{align}
&  \mid\Lambda_{N}(t)-\mid\gamma+t\mid^{2}\mid<\frac{1}{2}\rho^{\alpha_{1}},\\
&  \mid b(N,\gamma)\mid>c_{4}\rho^{-c\alpha}%
\end{align}
hold, \ where $b(N,\gamma)=(\Psi_{N,t},e^{i(\gamma+t,x)}),$ $\Psi_{N,t}(x)$ is
a normalized eigenfunction of $L_{t}(q(x))$ corresponding to $\Lambda_{N}(t),$
then $\ $the following statements are valid:

(a) if $\gamma+t$ is in the non-resonance domain, then $\Lambda_{N}(t)$
satisfies (5) for $k=1,2,...,[\frac{1}{3}(p-c)]$ ( see Theorem 1);

(b) if $\gamma+t\in E_{s}\backslash E_{s+1},$ where $s=1,2,...,d-1,$ then%
\begin{equation}
\Lambda_{N}(t)=\lambda_{j}(\gamma+t)+O(\mid\gamma+t\mid^{-k\alpha}),
\end{equation}
where $\lambda_{j}$ is an eigenvalue of the matrix $C(\gamma+t)$ ( see (26)
and Theorem 2). Moreover, we prove that every big eigenvalue of the operator
$L_{t}(q(x))$ for all values of quasimomenta $t$ satisfies one of these formulae.

For investigation of the Bloch function, in section 3, we find the values of
quasimomenta $\gamma+t$ for which the corresponding eigenvalues are simple ,
namely we construct the subset $B$ of $U(\rho^{\alpha_{1}},p)$ with the
following properties:

Pr.1. If $\gamma+t\in B,$ then there exists a unique eigenvalue, denoted by
$\Lambda(\gamma+t),$ of the operator $L_{t}(q(x))$ satisfying (5). This is a
simple eigenvalue of $L_{t}(q(x))$. Therefore we call the set $B$ the simple
set of quasimomenta.

Pr.2. The eigenfunction $\Psi_{N(\gamma+t)}(x)\equiv\Psi_{\gamma+t}(x)$
corresponding to the eigenvalue $\Lambda(\gamma+t)$ is close to $e^{i(\gamma
+t,x)}$, namely
\begin{equation}
\Psi_{N}(x)=e^{i(\gamma+t,x)}+O(\mid\gamma+t\mid^{-\alpha_{1}}),
\end{equation}%
\begin{equation}
\Psi_{\gamma+t}(x)=e^{i(\gamma+t,x)}+\Phi_{k-1}(x)+O(\mid\gamma+t\mid
^{-k\alpha_{1}}),\text{ }k=1,2,...\text{ ,}%
\end{equation}
where $\Phi_{k-1}$ is explicitly expresed by $q(x)$ and the eigenvalues of
$L_{t}(0).$

Pr.3. The set $B$ contains the intervals $\{a+sb:s\in\lbrack-1,1]\}$ such that
$\Lambda(a-b)<\rho^{2},$ $\Lambda(a+b)>\rho^{2},$\ and $\Lambda(\gamma+t)$
\ is continuous on these intervals. Hence there exists $\gamma+t$ such that
$\Lambda(\gamma+t)=\rho^{2}$ for $\rho\gg1.$ It implies that there exist only
a finite number of gaps in the spectrum of $L,$ that is, it implies the
validity of Bethe-Sommerfeld conjecture for arbitrary dimension and for
arbitrary lattice.

Construction of the set $B$ consists of two steps.

Step 1. We prove that all eigenvalues $\Lambda_{N}(t)\sim\rho^{2}$ of the
operator $L_{t}(q(x))$ lie in the $\varepsilon_{1}=\rho^{-d-2\alpha}$
neighborhood of the numbers

$F(\gamma+t)=\mid\gamma+t\mid^{2}+F_{k_{1}-1}(\gamma+t)$, $\lambda_{j}%
(\gamma+t)$ ( see (5), (8)), where $k_{1}=[\frac{d}{3\alpha}]+2.$ We call
these numbers as the known parts of the eigenvalues. Moreover, for
$\gamma+t\in U(\rho^{\alpha_{1}},p)$ there is $\Lambda_{N}(t)$ satisfying
$\Lambda_{N}(t)=F(\gamma+t)+o(\varepsilon_{1})$ ( see (5))

Step 2. By eliminating the set of quasimomenta $\gamma+t$, for which the known
parts $F(\gamma+t)$ of $\Lambda_{N}(t)$ are situated from the known parts
$F(\gamma^{^{\prime}}+t),$ $\lambda_{j}(\gamma^{^{\prime}}+t)$ ($\gamma
^{^{\prime}}\neq\gamma)$ of other eigenvalues at a distance less than
$2\varepsilon_{1},$ we construct the set $B$ with the following properties: if
$\gamma+t\in B,$ then the following conditions (called simplicity conditions
for $\Lambda_{N}(t))$ hold
\begin{equation}
\mid F(\gamma+t)-F(\gamma^{^{\prime}}+t)\mid\geq2\varepsilon_{1}\text{ }%
\end{equation}
for $\gamma^{^{\prime}}\in K\backslash\{\gamma\},$ $\gamma^{^{\prime}}+t\in
U(\rho^{\alpha_{1}},p)$ and%
\begin{equation}
\mid F(\gamma+t)-\lambda_{j}(\gamma^{^{\prime}}+t)\mid\geq2\varepsilon_{1}%
\end{equation}
for $\gamma^{^{\prime}}\in K,\gamma^{^{\prime}}+t\in E_{k}\backslash
E_{k+1},j=1,2,...,$ where $K$ is the set of $\gamma^{^{\prime}}\in\Gamma$
satisfying $\mid F(\gamma+t)-\mid\gamma^{^{\prime}}+t\mid^{2}\mid<\frac{1}%
{3}\rho^{\alpha_{1}}$. Thus $B$ is the set of $\gamma+t\in U(\rho^{\alpha_{1}%
},p)$ satisfying the simplicity conditions (11), (12). As a consequence of
these conditions the eigenvalue $\Lambda_{N}(t)$ does not coincide with other
eigenvalues. To prove this, namely to prove the Pr.1 and (9), we show that for
any normalized eigenfunction $\Psi_{N}(x)$ corresponding to $\Lambda_{N}(t)$
the following equality holds:
\begin{equation}
\sum_{\gamma^{^{\prime}}\in\Gamma\backslash\gamma}\mid b(N,\gamma^{^{\prime}%
})\mid^{2}=O(\rho^{-2\alpha_{1}}).
\end{equation}

For the first time in [15-17] we constructed the simple set $B$ with the Pr.1
and Pr.3., though in those papers we emphasized the Bethe-Zommerfeld
conjecture. Note that for this conjecture and for Pr.1, Pr.3. it is enough to
prove that the left-hand side of (13) is less than $\frac{1}{4}$ ( we proved
this inequality in [15-17] and as noted in Theorem 3 of [16] and in [18] the
proof of this inequality does not differ from the proof of (13)). From (9) we
got \ (10) (see [18]) . But in those papers these results are written briefly.
The enlarged variant is written in [19] which can not be used as reference. In
this paper we write these results in improved and enlarged form. The main
difficulty and the crucial point of papers [15-17] were the construction of
the simple set $B$ with the Pr.1.,Pr.3. This difficulty of the perturbation
theory of $L(q(x))$ is of a physical nature and it is connected with the
complicated picture of the chystal diffraction. If $d=2,3,$ then
$F(\gamma+t)=\mid\gamma+t\mid^{2}$ and the matrix $C(\gamma+t)$ corresponds to
the Schrodinger operator with directional potential $q_{\delta}(x)=\sum_{n\in
Z}q_{n\delta}e^{in(\delta,x)}$ ( see [16]). So for construction of the simple
set $B$ of quasimomenta we eliminated the vicinities of the diffraction planes
and the sets connected with directional potential ( see (11), (12)).
Besides,\ for nonsmooth potentials $q(x)\in L_{2}(\mathbb{R}^{2}/\Omega),$ we
eliminated a set, which is described in the terms of the number of states (
see [15,19]). The simple sets $B$ of quasimomenta for the first time is
constructed and investigated ( hence the main difficulty and the crucial point
of perturbation theory of $L(q)$ is investigated) in [16] for $d=3$ and in
[15,17] for the cases:

1. $d=2,$ $q(x)\in L_{2}(F);$ \ \ \ 2. $d>2,$ $q(x)$ is a smooth potential.

Then, Yu.E. Karpeshina proved ( see [7-9]) the convergence of the perturbation
series of two and three dimensional Schrodinger operator $L(q)$ with a wide
class of nonsmooth potential $q(x)$ for a set, that is similar to $B$, of
quasimomenta. In papers [3,4] the asymptotic formulas for the eigenvalues and
Bloch function of the two and three dimensional operator $L_{t}(q(x))$ were
obtained. In [5] the asymptotic formulae for the eigenvalues of $L_{0}(q(x))$
were obtained.

In section 4 we consider the geometrical aspects of the simple sets. We prove
that the simple sets $B$ has asymptotically full measure on $\mathbb{R}^{d}$.
Moreover we construct a part of isoenergetic surfaces corresponding to
$\rho^{2},$ which is smooth surfaces and has the measure asymptotically close
to the measure of the Fermi surfaces $\{x\in R:\mid x\mid=\rho\}$ of the
operator $L(0).$ The nonemptyness of the Fermi surfaces \ for $\rho\gg1$
implies the the validity of the Bethe-Sommerfeld conjecture.

For the first time M.M. Skriganov [11,12] proved the validity of the
Bethe-Sommerfeld conjecture for the Scrodinger operator for dimension $d=2,3$
for arbitrary lattice, for dimension $d>3$ for rational lattice. The
Skriganov's method is based on the detail investigation of the arithmetic and
geometric properties of the lattice. B.E.J.Dahlberg \ and E.Trubowits [1]
using an asymptotic of Bessel function, gave the simple proof of this
conjecture for the two dimensional Scrodinger operator. Then in papers [15-17]
we proved the validity of the Bethe-Sommerfeld conjecture for arbitrary
lattice and for arbitrary dimension by using the asymptotic formulas and by
construction of the simple set $B,$ that is, by\ the method of perturbation
theory. Yu.E. Karpeshina ( see [7-9]) proved this conjecture for two and three
dimensional Schrodinger operator $L(q)$ for a wide class of singular
potentials $q(x),$ including Coulomp potential, by \ the method of
perturbation theory. B. Helffer and A. Mohamed [6], by investigations the
integrated density of states, proved the validity of the Bethe-Sommerfeld
conjecture for the Scrodinger operator for $d\leq4$ for arbitrary lattice.
Recently L. Parnovski and A. V. Sobelev [10] proved this conjecture for
$d\leq4.$ The method of this paper and papers [15-17] is a first and uniqie,
for the present, by which the validity of the Bethe-Sommerfeld conjecture for
arbitrary lattice and for arbitrary dimension is proved.

In this paper for the different types of the measures of the subset $A$ of
$\mathbb{R}^{d}$ we use the same notation $\mu(A).$ By $\mid A\mid$ we denote
the number of elements of the set $A\subset\Gamma$ and use the following
obvious fact. If $a\sim\rho,$ then the number of elements of the set
$\{\gamma+t:$ $\gamma\in\Gamma\}$ satisfying $\mid\mid\gamma+t\mid-a\mid<1$ is
less than $c_{5}\rho^{d-1}.$ Therefore the number of eigenvalues of $L_{t}(q)$
lying in $(a^{2}-\rho,a^{2}+\rho)$ is less than $c_{5}\rho^{d-1}.$ Besides, we
use the inequalities:%
\begin{align}
\alpha_{1}+d\alpha &  <1-\alpha\,,\ \ \ \ \ d\alpha<\frac{1}{2}\alpha
_{d},\ \ \ k_{1}\leq\frac{1}{3}(p-\frac{1}{2}(q(d-1)),\\
p_{1}\alpha_{1}  &  \geq p\alpha,\ \ \ \ \ 3k_{1}\alpha>d+2\alpha
,\ \ \ \ \ \ \alpha_{k}+(k-1)\alpha<1,\nonumber\\
\alpha_{k+1}  &  >2(\alpha_{k}+(k-1))\alpha\nonumber
\end{align}
for $k=1,2,...,d,$ which follow from the definitions $p=s-d,$ $\alpha
_{k}=3^{k}\alpha,$ $\alpha=\frac{1}{q},$ $q=3^{d}+d+2,$ $k_{1}=[\frac
{d}{3\alpha}]+2,$ $p_{1}=[\frac{p}{3}]+1$ of the numbers $p,q,\alpha
_{k},\alpha,k_{1},p_{1}.$

\section{Asymptotic Formulae for Eigenvalues}

First we obtain the asymptotic formulas for the non-resonance eigenvalues by
iteration of the formula%
\begin{equation}
(\Lambda_{N}-\mid\gamma+t\mid^{2})b(N,\gamma)=(\Psi_{N,t}(x)q(x),e^{i(\gamma
+t,x)}),
\end{equation}
which is obtained from equation $-\Delta\Psi_{N,t}(x)+q(x)\Psi_{N,t}%
(x)=\Lambda_{N}\Psi_{N,t}(x)$ by multiplying by $e^{i(\gamma+t,x)}).$
Introducing into (15) the expansion (3) of $q(x)$, we get
\begin{equation}
(\Lambda_{N}-\mid\gamma+t\mid^{2})b(N,\gamma)=\sum_{\gamma_{1}\in\Gamma
(\rho^{\alpha})}q_{\gamma_{1}}b(N,\gamma-\gamma_{1})+O(\rho^{-p\alpha}).
\end{equation}
From the relations (15), (16) it follows that
\begin{equation}
b(N,\gamma^{^{\prime}})=\dfrac{(\Psi_{N,t}q(x),e^{i(\gamma^{^{\prime}}+t,x)}%
)}{\Lambda_{N}-\mid\gamma^{^{\prime}}+t\mid^{2}}=%
{\displaystyle\sum_{\gamma_{1}\in\Gamma(\rho^{\alpha})}}
\dfrac{q_{\gamma_{1}}b(N,\gamma^{^{\prime}}-\gamma_{1})}{\Lambda_{N}%
-\mid\gamma^{^{\prime}}+t\mid^{2}}+O(\rho^{-p\alpha})
\end{equation}
for all vectors $\gamma^{^{\prime}}\in\Gamma$ satisfying the inequality
\begin{equation}
\mid\Lambda_{N}-\mid\gamma^{^{\prime}}+t\mid^{2}\mid>\frac{1}{2}\rho
^{\alpha_{1}}.
\end{equation}
This inequality is called the iterability condition. If (6) holds and
$\mid\gamma+t\mid^{2}$ is a non-resonance eigenvalue, i.e., $\gamma+t\in
U(\rho^{\alpha_{1}},p),$ then
\begin{equation}
\mid\mid\gamma+t\mid^{2}-\mid\gamma-\gamma_{1}+t\mid^{2}\mid>\rho^{\alpha_{1}%
},\text{ }\mid\Lambda_{N}-\mid\gamma-\gamma_{1}+t\mid^{2}\mid>\frac{1}{2}%
\rho^{\alpha_{1}}\text{ }%
\end{equation}
for all $\gamma_{1}\in\Gamma(p\rho^{\alpha}).$ Hence the vector $\gamma
-\gamma_{1}$ for $\gamma+t\in U(\rho^{\alpha_{1}},p)$ and $\gamma_{1}\in
\Gamma(p\rho^{\alpha})$ satisfies (18). Therefore, in (17) one can replace
$\gamma^{^{\prime}}$ by $\gamma-\gamma_{1}$ and write
\[
b(N,\gamma-\gamma_{1})=%
{\displaystyle\sum_{\gamma_{2}\in\Gamma(\rho^{\alpha})}}
\dfrac{q_{\gamma_{2}}b(N,\gamma-\gamma_{1}-\gamma_{2})}{\Lambda_{N}-\mid
\gamma-\gamma_{1}+t\mid^{2}}+O(\rho^{-p\alpha}).
\]
Substituting this for $b(N,\gamma-\gamma_{1})$ into right-hand side of (16)
and isolating the terms containing the multiplicand $b(N,\gamma)$, we get
\[
(\Lambda_{N}-\mid\gamma+t\mid^{2})b(N,\gamma)=%
{\displaystyle\sum_{\gamma_{1},\gamma_{2}\in\Gamma(\rho^{\alpha})}}
\dfrac{q_{\gamma_{1}}q_{\gamma_{2}}b(N,\gamma-\gamma_{1}-\gamma_{2})}%
{\Lambda_{N}-\mid\gamma-\gamma_{1}+t\mid^{2}}+O(\rho^{-p\alpha})=
\]%
\[%
{\displaystyle\sum_{\gamma_{1}\in\Gamma(\rho^{\alpha})}}
\dfrac{\mid q_{\gamma_{1}}\mid^{2}b(N,\gamma)}{\Lambda_{N}-\mid\gamma
-\gamma_{1}+t\mid^{2}}+%
{\displaystyle\sum_{\substack{\gamma_{1},\gamma_{2}\in\Gamma(\rho^{\alpha
}),\\\gamma_{1}+\gamma_{2}\neq0}}}
\dfrac{q_{\gamma_{1}}q_{\gamma_{2}}b(N,\gamma-\gamma_{1}-\gamma_{2})}%
{\Lambda_{N}-\mid\gamma-\gamma_{1}+t\mid^{2}}+O(\rho^{-p\alpha}),
\]
since $q_{\gamma_{1}}q_{\gamma_{2}}=\mid q_{\gamma_{1}}\mid^{2}$ for
$\gamma_{1}+\gamma_{2}=0$ and the last sum is taken under the condition
$\gamma_{1}+\gamma_{2}\neq0.$ Repeating this process $p_{1}\equiv\lbrack
\frac{p}{3}]+1$ times, i.e., in the last formula replacing $b(N,\gamma
-\gamma_{1}-\gamma_{2})$ by its expression from (17) ( in (17) replace
$\gamma^{^{\prime}}$ by $\gamma-\gamma_{1}-\gamma_{2}$) and isolating the
terms containing $b(N,\gamma)$ etc., we obtain
\begin{equation}
(\Lambda_{N}-\mid\gamma+t\mid^{2})b(N,\gamma)=A_{p_{1}}(\Lambda_{N}%
,\gamma+t)b(N,\gamma)+C_{p_{1}}+O(\rho^{-p\alpha}),
\end{equation}
where $A_{p_{1}}(\Lambda_{N},\gamma+t)=\sum_{k=1}^{p_{1}}S_{k}(\Lambda
_{N},\gamma+t)$ ,
\[
S_{k}(\Lambda_{N},\gamma+t)=%
{\displaystyle\sum_{\gamma_{1},...,\gamma_{k}\in\Gamma(\rho^{\alpha})}}
\dfrac{q_{\gamma_{1}}q_{\gamma_{2}}...q_{\gamma_{k}}q_{-\gamma_{1}-\gamma
_{2}-...-\gamma_{k}}}{\prod_{j=1}^{k}(\Lambda_{N}-\mid\gamma+t-\sum_{i=1}%
^{j}\gamma_{i}\mid^{2})},
\]%
\[
C_{p_{1}}=\sum_{\gamma_{1},...,\gamma_{p_{1}+1}\in\Gamma(\rho^{\alpha})}%
\dfrac{q_{\gamma_{1}}q_{\gamma_{2}}...q_{\gamma_{p_{1}+1}}b(N,\gamma
-\gamma_{1}-\gamma_{2}-...-\gamma_{p_{1}+1})}{\prod_{j=1}^{p_{1}}(\Lambda
_{N}-\mid\gamma+t-\sum_{i=1}^{j}\gamma_{i}\mid^{2})}.
\]
Here the sums for $S_{k}$ and $C_{p_{1}}$ are taken under the additional conditions

$\gamma_{1}+\gamma_{2}+...+\gamma_{s}\neq0$ for $s=1,2,...,k$ and
$s=1,2,...,p_{1}$ respectively. These conditions and the inclusion $\gamma
_{i}\in\Gamma(\rho^{\alpha})$ for $i=1,2,...,p_{1}$ imply the relation
$\sum_{i=1}^{j}\gamma_{i}\in\Gamma(p\rho^{\alpha})$. Therefore from the second
inequality in (19) it follows that the absolute values of the denominators of
the fractions in $S_{k}$ and $C_{p_{1}}$ are greater than $(\frac{1}{2}%
\rho^{\alpha_{1}})^{k}$ and $(\frac{1}{2}\rho^{\alpha_{1}})^{p_{1}}$
respectively. Hence the first inequality in (4) and $p_{1}\alpha_{1}\geq
p\alpha$ ( see the fourth inequality in (14)) yield
\begin{equation}
C_{p_{1}}=O(\rho^{-p_{1}\alpha_{1}})=O(\rho^{-p\alpha}),\text{ }S_{k}%
(\Lambda_{N},\gamma+t)=O(\rho^{-k\alpha_{1}}),\forall k=1,2,...,p_{1}.
\end{equation}
Since we used only the condition (6) for $\Lambda_{N},$ it follows that
\begin{equation}
S_{k}(a,\gamma+t)=O(\rho^{-k\alpha_{1}})
\end{equation}
for all $a\in\mathbb{R}$ satisfying $\mid a-\mid\gamma+t\mid^{2}\mid<\frac
{1}{2}\rho^{\alpha_{1}}.$ Thus finding $N$ such that $\Lambda_{N}$ is close to
$\mid\gamma+t\mid^{2}$ and $b(N,\gamma)$ is not very small, then dividing both
sides of (20) by $b(N,\gamma),$ we get the asymptotic formulas for
$\Lambda_{N}$.

\begin{theorem}
$(a)$ Suppose $\gamma+t\in U(\rho^{\alpha_{1}},p),$ $\mid\gamma\mid\sim\rho.$
If (6) and (7) hold, then $\Lambda_{N}$ satisfies formulas (5) for
$k=1,2,...,[\frac{1}{3}(p-c)],$ where
\begin{equation}
F_{s}=O(\rho^{-\alpha_{1}}),\forall s=0,1,...,
\end{equation}
and $F_{0}=0,$ $F_{s}=A_{s}(\mid\gamma+t\mid^{2}+F_{s-1},\gamma+t)$ for
$s=1,2,....$

$(b)$ For $\gamma+t\in U(\rho^{\alpha_{1}},p),$ $\mid\gamma\mid\sim\rho$ there
exists an eigenvalue $\Lambda_{N}$ of $L_{t}(q(x))$ satisfying (5).
\end{theorem}

\begin{proof}
$(a)$ To prove (5) in case $k=1$ we divide both side of (20) by $b(N,\gamma)$
and use (7), (21). Then we obtain
\begin{equation}
\Lambda_{N}-\mid\gamma+t\mid^{2}=\mid\gamma+t\mid^{2}+O(\rho^{-\alpha_{1}}).
\end{equation}
This and $\alpha_{1}=3\alpha$ ( see the end of the introduction) imply that
formula (5) for $k=1$ holds and $F_{0}=0.$ Hence (23) for $s=0$ is also
proved. Moreover, from (22), we obtain $S_{k}(\mid\gamma+t\mid^{2}%
+O(\rho^{-\alpha_{1}}),\gamma+t)=O(\rho^{-\alpha_{1}})$ for $k=1,2,....$
Therefore (23) for arbitrary $s$ follows from the definition of $F_{s}$ by
induction. Now we prove (5) by induction on $k$. Suppose (5) holds for $k=j$,
that is,

$\Lambda_{N}=\mid\gamma+t\mid^{2}+F_{k-1}(\gamma+t)+O(\rho^{-3k\alpha}).$
Substituting this into $A_{p_{1}}(\Lambda_{N},\gamma+t)$ in (20) and dividing
both sides of (20) by $b(N,\gamma),$ we get
\begin{align*}
\Lambda_{N}  &  =\mid\gamma+t\mid^{2}+A_{p_{1}}(\mid\gamma+t\mid^{2}%
+F_{j-1}+O(\rho^{-j\alpha_{1}}),\gamma+t)+O(\rho^{-(p-c)\alpha})=\\
&  \mid\gamma+t\mid^{2}+\{A_{p_{1}}(\mid\gamma+t\mid^{2}+F_{j-1}%
+O(\rho^{-j\alpha_{1}}),\gamma+t)-\\
A_{p_{1}}(  &  \mid\gamma+t\mid^{2}+F_{j-1},\gamma+t)\}+A_{p_{1}}(\mid
\gamma+t\mid^{2}+F_{j-1},\gamma+t)+O(\rho^{-(p-c)\alpha}).
\end{align*}
To prove $(a)$ for $k=j+1$ we need to show that the expression in curly
brackets is equal to $O(\rho^{-(j+1)\alpha_{1}}).$ It can be checked by using
(4), (19), (23) and the obvious relation
\begin{align*}
&  \frac{1}{\prod_{j=1}^{s}(\mid\gamma+t\mid^{2}+F_{j-1}+O(\rho^{-j\alpha_{1}%
})-\mid\gamma+t-\sum_{i=1}^{s}\gamma_{i}\mid^{2})}-\\
&  \frac{1}{\prod_{j=1}^{s}(\mid\gamma+t\mid^{2}+F_{j-1}-\mid\gamma
+t-\sum_{i=1}^{s}\gamma_{i}\mid^{2})}\\
&  =\frac{1}{\prod_{j=1}^{s}(\mid\gamma+t\mid^{2}+F_{j-1}-\mid\gamma
+t-\sum_{i=1}^{s}\gamma_{i}\mid^{2})}(\frac{1}{1-O(\rho^{-(j+1)\alpha_{1}}%
)}-1)
\end{align*}
$=O(\rho^{-(j+1)\alpha_{1}})$ for $s=1,2,...,p_{1}.$

$(b)$ Let $A$ be the set of indices $N$ satisfying (6). Using (15) and Bessel
inequality, we obtain%
\[
\sum_{N\notin A}\mid b(N,\gamma)\mid^{2}=\sum_{N\notin A}\mid\dfrac{(\Psi
_{N}(x),q(x)e^{i(\gamma+t,x)})}{\Lambda_{N}-\mid\gamma+t\mid^{2}}\mid
^{2}=O(\rho^{-2\alpha_{1}})
\]
Hence, by the Parseval equality, we have $\sum_{N\in A}\mid b(N,\gamma
)\mid^{2}=1-O(\rho^{-2\alpha_{1}}).$ This and the inequality $\mid A\mid
<c_{5}\rho^{d-1}=c_{5}\rho^{(d-1)q\alpha}$ ( see the end of the introduction)
imply that there exists a number $N$ satisfying $\mid b(N,\gamma)\mid>\frac
{1}{2}(c_{5})^{-1}\rho^{-\frac{(d-1)q}{2}\alpha}$, that is, (7) holds for
$c=\frac{(d-1)q}{2}$ . Thus $\Lambda_{N}$ satisfies (5) due to $(a)$
\end{proof}

Theorem 1 shows that in the non-resonance case the eigenvalue of the perturbed
operator $L_{t}(q(x))$ is close to the eigenvalue of the unperturbed operator
$L_{t}(0).$ However, in theorem 2 we prove that if $\gamma+t\in\cap_{i=1}%
^{k}V_{\gamma_{i}}(\rho^{\alpha_{k}})\backslash E_{k+1}$ for $k\geq1,$ where
$\gamma_{1},\gamma_{2},...,\gamma_{k}$ are linearly independent vectors of
$\Gamma(p\rho^{\alpha}),$ then the corresponding eigenvalue of $L_{t}(q(x))$
is close to the eigenvalue of the matrix constructed as follows. Introduce the sets:

$B_{k}\equiv B_{k}(\gamma_{1},\gamma_{2},...,\gamma_{k})=\{b:b=\sum_{i=1}%
^{k}n_{i}\gamma_{i},n_{i}\in Z,\mid b\mid<\frac{1}{2}\rho^{\frac{1}{2}%
\alpha_{k+1}}\},$%
\begin{equation}
B_{k}(\gamma+t)=\gamma+t+B_{k}=\{\gamma+t+b:b\in B_{k}\},
\end{equation}
$B_{k}(\gamma+t,p_{1})=\{\gamma+t+b+a:b\in B_{k},\mid a\mid<p_{1}\rho^{\alpha
},a\in\Gamma\}.$

Denote by $h_{i}+t$ for $i=1,2,...,b_{k}$ the vectors of $B_{k}(\gamma
+t,p_{1}),$ where

$b_{k}\equiv b_{k}(\gamma_{1},\gamma_{2},...,\gamma_{k})$ is the number of the
vectors of $B_{k}(\gamma+t,p_{1})$. Define the matrix $C(\gamma+t,\gamma
_{1},\gamma_{2},...,\gamma_{k})\equiv(c_{i,j})$ by the formulas
\begin{equation}
c_{i,i}=\mid h_{i}+t\mid^{2},\text{ }c_{i,j}=q_{h_{i}-h_{j}},\text{ }\forall
i\neq j,
\end{equation}
where $i,j=1,2,...,b_{k}.$ We consider the resonance eigenvalue \ $\mid
\gamma+t\mid^{2}$ for $\gamma+t\in(\cap_{i=1}^{k}V_{\gamma_{i}}(\rho
^{\alpha_{k}}))$ by using the following lemma.

\begin{lemma}
If $\gamma+t\in\cap_{i=1}^{k}V_{\gamma_{i}}(\rho^{\alpha_{k}})\backslash
E_{k+1},$ $h+t\in B_{k}(\gamma+t,p_{1}),$

$(h-\gamma^{^{\prime}}+t)\notin B_{k}(\gamma+t,p_{1}),$ then
\begin{equation}
\mid\mid\gamma+t\mid^{2}-\mid h-\gamma^{^{\prime}}-\gamma_{1}^{^{\prime}%
}-\gamma_{2}^{^{\prime}}-...-\gamma_{s}^{^{\prime}}+t\mid^{2}\mid>\frac{1}%
{5}\rho^{\alpha_{k+1}},
\end{equation}
where $\gamma^{^{\prime}}\in\Gamma(\rho^{\alpha}),$ $\gamma_{j}^{^{\prime}}%
\in\Gamma(\rho^{\alpha}),$ $j=1,2,...,s$ and $s=0,1,...,p_{1}-1.$
\end{lemma}

\begin{proof}
The inequality $p>2p_{1}$ ( see the end of the introduction) and the
conditions of the lemma 1 imply that

$h-\gamma^{^{\prime}}-\gamma_{1}^{^{\prime}}-\gamma_{2}^{^{\prime}}%
-...-\gamma_{s}^{^{\prime}}+t\in B_{k}(\gamma+t,p)\backslash B_{k}(\gamma+t)$
for all $s=0,1,...,p_{1}-1.$ It follows from the definitions of $B_{k}%
(\gamma+t,p),B_{k}$ that ( see (25))

$h-\gamma^{^{\prime}}-\gamma_{1}^{^{\prime}}-\gamma_{2}^{^{\prime}}%
-...-\gamma_{s}^{^{\prime}}+t=\gamma+t+b+a,$ where
\begin{equation}
\mid b\mid<\frac{1}{2}\rho^{\frac{1}{2}\alpha_{k+1}},\mid a\mid<p\rho^{\alpha
},\text{ }\gamma+t+b+a\notin\gamma+t+B_{k}.
\end{equation}
Then (27) has the form
\begin{equation}
\mid\mid\gamma+t+a+b\mid^{2}-\mid\gamma+t\mid^{2}\mid>\frac{1}{5}\rho
^{\alpha_{k+1}}.
\end{equation}
To prove (29) we consider two cases:

Case 1. $a\in P$, where $P=Span\{\gamma_{1,}\gamma_{2},...,\gamma_{k}\}.$
Since $b\in B_{k}\subset P,$ we have $a+b\in P.$ This with the third relation
in (28) imply that $a+b\in P\backslash B_{k}$ ,i.e., $\mid a+b\mid\geq\frac
{1}{2}$ $\rho^{\frac{1}{2}\alpha_{k+1}}$. Consider the orthogonal
decomposition $\gamma+t=y+v$ of $\gamma+t,$ where $v\in P$ and $y\bot P.$
First we prove that the projection $v$ of any vector $x\in\cap_{i=1}%
^{k}V_{\gamma_{i}}(\rho^{\alpha_{k}})$ on $P$ satisfies
\begin{equation}
\mid v\mid=O(\rho^{(k-1)\alpha+\alpha_{k}}).
\end{equation}
For this we turn the coordinate axis so that $Span\{\gamma_{1,}\gamma
_{2},...,\gamma_{k}\}$ coincides with the span of the vectors $e_{1}%
=(1,0,0,...,0)$, $e_{2}=(0,1,0,...,0),...,$ $e_{k}$. Then $\gamma_{s}%
=\sum_{i=1}^{k}\gamma_{s,i}e_{i}$ for $s=1,2,...,k$ . Therefore the relation
$x\in\cap_{i=1}^{k}V_{\gamma_{i}}(\rho^{\alpha_{k}})$ implies that
\[
\sum_{i=1}^{k}\gamma_{s,i}x_{i}=O(\rho^{\alpha_{k}}),s=1,2,...,k;\text{ }%
x_{n}=\frac{\det(b_{j,i}^{n})}{\det(\gamma_{j,i})}\text{, }n=1,2,...,k,
\]
where $x=(x_{1},x_{2},...,x_{d}),\gamma_{j}=(\gamma_{j,1},\gamma
_{j,2},...,\gamma_{j,k},0,0,...,0),$ $b_{j,i}^{n}=\gamma_{j,i}$ for $n\neq j$
and $b_{j,i}^{n}=O(\rho^{\alpha_{k}})$ for $n=j.$ Taking into account that the
determinant $\det(\gamma_{j,i})$ is a volume of the parallelepiped
$\{\sum_{i=1}^{k}b_{i}\gamma_{i}:b_{i}\in\lbrack0,1],i=1,2,...,k\}$ and using
$\mid\gamma_{j,i}\mid<p\rho^{\alpha}$ ( since $\gamma_{j}\in\Gamma
(p\rho^{\alpha})$ ), we get the estimations
\begin{equation}
x_{n}=O(\rho^{\alpha_{k}+(k-1)\alpha})\text{ ,}\forall n=1,2,...,k;\text{
}\forall x\in\cap_{i=1}^{k}V_{\gamma_{i}}(\rho^{\alpha_{k}}).
\end{equation}
Hence (30) holds. Therefore, using the inequalities $\mid a+b\mid\geq\frac
{1}{2}$ $\rho^{\frac{1}{2}\alpha_{k+1}}$ ( see above), $\alpha_{k+1}%
>2(\alpha_{k}+(k-1)\alpha)$ ( see the \ seventh inequality in (14)), and the
obvious equalities $(y,v)=(y,a)=(y,b)=0,$%
\begin{equation}
\mid\gamma+t+a+b\mid^{2}-\mid\gamma+t\mid^{2}=\mid a+b+v\mid^{2}-\mid
v\mid^{2},
\end{equation}
we obtain the estimation (29).

Case 2. $a\notin P.$ First we show that
\begin{equation}
\mid\mid\gamma+t+a\mid^{2}-\mid\gamma+t\mid^{2}\mid\geq\rho^{\alpha_{k+1}}.
\end{equation}
Suppose, to the contrary, that it does not hold. Then $\gamma+t\in V_{a}%
(\rho^{\alpha_{k+1}}).$ On the other hand $\gamma+t\in\cap_{i=1}^{k}%
V_{\gamma_{i}}(\rho^{\alpha_{k+1}})$ ( see the conditions of Lemma 1).
Therefore we have $\gamma+t\in E_{k+1}$ which contradicts the conditions of
the lemma. \ So (33) is proved. Now, to prove (29) we write the difference
$\mid\gamma+t+a+b\mid^{2}-\mid\gamma+t\mid^{2}$ as the sum of $d_{1}\equiv
\mid\gamma+t+a+b\mid^{2}-\mid\gamma+t+b\mid^{2}$ and $d_{2}\equiv\mid
\gamma+t+b\mid^{2}-\mid\gamma+t\mid^{2}.$ Since $d_{1}=\mid\gamma+t+a\mid
^{2}-\mid\gamma+t\mid^{2}+2(a,b),$ it follows from the inequalities (33), (28)
that $\mid d_{1}\mid>\frac{2}{3}$ $\rho^{\alpha_{k+1}}$. On the other hand,
taking $a=0$ in (32) we have $d_{2}=\mid b+v\mid^{2}-\mid v\mid^{2}.$
Therefore (30), the first inequality in (28) and the \ seventh inequality in
(14) imply that $\mid d_{2}\mid<\frac{1}{3}$ $\rho^{\alpha_{k+1}},$ $\mid
d_{1}\mid-\mid d_{2}\mid>\frac{1}{3}\rho^{\alpha_{k+1}},$ that is, (29) holds
\end{proof}

\begin{theorem}
$(a)$ Suppose $\mid\gamma\mid\sim\rho,$ $\gamma+t\in(\cap_{i=1}^{k}%
V_{\gamma_{i}}(\rho^{\alpha_{k}}))\backslash E_{k+1},$ where

$k=1,2,...,d-1.$ If (6) and (7) hold, then there is an index $j$ such that
\begin{equation}
\Lambda_{N}=\lambda_{j}(\gamma+t)+O(\rho^{-(p-c-\frac{1}{4}d3^{d})\alpha}),
\end{equation}
where $\lambda_{1}(\gamma+t)\leq\lambda_{2}(\gamma+t)\leq...\leq\lambda
_{b_{k}}(\gamma+t)$ are the eigenvalues of the matrix $C(\gamma+t,\gamma
_{1},\gamma_{2},...,\gamma_{k})$ defined in (26).

$(b)$ Every eigenvalue $\Lambda_{N}(t)\sim\rho^{2}$ of the operator
$L_{t}(q(x))$ satisfies either (5) or (34) for $c=\frac{q(d-1)}{2}.$
\end{theorem}

\begin{proof}
$(a)$Writing the equation (16) for all $h_{i}+t\in B_{k}(\gamma+t,p_{1}),$ we
obtain%
\begin{equation}
(\Lambda_{N}-\mid h_{i}+t\mid^{2})b(N,h_{i})=\sum_{\gamma^{^{\prime}}\in
\Gamma(\rho^{\alpha})}q_{\gamma^{^{\prime}}}b(N,h_{i}-\gamma^{^{\prime}%
})+O(\rho^{-p\alpha})
\end{equation}
for $i=1,2,...,b_{k}$ ( see (25) for definition of $B_{k}(\gamma+t,p_{1})$).
It follows from (6) and lemma 1 that if $(h_{i}-\gamma^{^{\prime}}+t)\notin
B_{k}(\gamma+t,p_{1}),$ then
\[
\mid\Lambda_{N}-\mid h_{i}-\gamma^{^{\prime}}-\gamma_{1}-\gamma_{2}%
-...-\gamma_{s}+t\mid^{2}\mid>\frac{1}{6}\rho^{\alpha_{k+1}},
\]
where $\gamma^{^{\prime}}\in\Gamma(\rho^{\alpha}),\gamma_{j}\in\Gamma
(\rho^{\alpha}),$ $j=1,2,...,s$ and $s=0,1,...,p_{1}-1.$ Therefore, applying
the formula (17) $p_{1}$ times, using (4) and $p_{1}\alpha_{k+1}>p_{1}%
\alpha_{1}\geq p\alpha$ ( see the\ fourth inequality in (14)), we see that if
$(h_{i}-\gamma^{^{\prime}}+t)\notin B_{k}(\gamma+t,p_{1}),$ then
\begin{equation}
b(N,h_{i}-\gamma^{^{\prime}})=\nonumber
\end{equation}%
\begin{equation}
\sum_{\gamma_{1},...,\gamma_{p_{1}-1}\in\Gamma(\rho^{\alpha})}\dfrac
{q_{\gamma_{1}}q_{\gamma_{2}}...q_{\gamma_{p_{1}}}b(N,h_{i}-\gamma^{^{\prime}%
}-\sum_{i=1}^{p_{1}}\gamma_{i})}{\prod_{j=0}^{p_{1}-1}(\Lambda_{N}-\mid
h_{i}-\gamma^{^{\prime}}+t-\sum_{i=1}^{j}\gamma_{i}\mid^{2})}+
\end{equation}%
\[
+O(\rho^{-p\alpha})=O(\rho^{p_{1}\alpha_{k+1}})+O(\rho^{-p\alpha}%
)=O(\rho^{-p\alpha}).
\]
Hence (35) has the form%
\[
(\Lambda_{N}-\mid h_{i}+t\mid^{2})b(N,h_{i})=\sum_{\gamma^{^{\prime}}%
}q_{\gamma^{^{\prime}}}b(N,h_{i}-\gamma^{^{\prime}})+O(\rho^{-p\alpha
}),i=1,2,...,b_{k},
\]
where the sum is taken under the conditions $\gamma^{^{\prime}}\in\Gamma
(\rho^{\alpha})$ and

$h_{i}-\gamma^{^{\prime}}+t\in B_{k}(\gamma+t,p_{1})$. It can be written in
matrix form
\[
(C-\Lambda_{N}I)(b(N,h_{1}),b(N,h_{2}),...b(N,h_{b_{k}}))=O(\rho^{-p\alpha}),
\]
where the rigth-hand side of this system is a vector having the norm

$\mid\mid O(\rho^{-p\alpha})\mid\mid=O(\sqrt{b_{k}}\rho^{-p\alpha})$. Now,
taking into account that

$\gamma+t\in\{h_{i}+t:i=1,2,...,b_{k}\}$ and (7) holds, we have
\begin{align}
c_{4}\rho^{-c\alpha}  &  <(\sum_{i=1}^{b_{k}}\mid b(N,h_{i})\mid^{2}%
)^{\frac{1}{2}}\leq\parallel(C-\Lambda_{N}I)^{-1}\parallel\sqrt{b_{k}}%
c_{6}\rho^{-p\alpha},\\
\max_{i=1,2,...,b_{k}}  &  \mid\Lambda_{N}-\lambda_{i}\mid^{-1}=\parallel
(C-\Lambda_{N}I)^{-1}\parallel>c_{4}c_{6}^{-1}b_{k}^{-\frac{1}{2}}%
\rho^{-c\alpha+p\alpha}.
\end{align}
Since $b_{k}$ is the number of the vectors of $B_{k}(\gamma+t,p_{1}),$ it
follows from the definition of $B_{k}(\gamma+t,p_{1})$ ( see (25)) and the
obvious relations$\mid B_{k}\mid=O(\rho^{\frac{k}{2}\alpha_{k+1}}),$
$\mid\Gamma(p_{1}\rho^{\alpha})\mid=O(\rho^{d\alpha})$ and $d\alpha<\frac
{1}{2}\alpha_{d}$ ( see the end of introduction), we get
\begin{equation}
b_{k}=O(\rho^{d\alpha+\frac{k}{2}\alpha_{k+1}})=O(\rho^{\frac{d}{2}\alpha_{d}%
})=O(\rho^{\frac{d}{2}3^{d}\alpha}),\forall k=1,2,...,d-1
\end{equation}
Thus formula (34) follows from (38) and (39).

$(b)$ Let $\Lambda_{N\text{ }}(t)$ be any eigenvalue of order $\rho^{2}$ of
the operator $L_{t}(q(x)).$ Denote by $D$ the set of all vectors $\gamma
\in\Gamma$ satisfying (6). From (15), arguing as in the proof of Theorem
1($b$), we obtain

$\sum_{\gamma\in D}\mid b(N,\gamma)\mid^{2}=1-O(\rho^{-2\alpha_{1}}).$ Since
$\mid D\mid=O(\rho^{d-1})$ ( see the end of the introduction), there exists
$\gamma\in D$ such that

$\mid b(N,\gamma)\mid>c_{7}\rho^{-\frac{(d-1)}{2}}=c_{7}\rho^{-\frac
{(d-1)q}{2}\alpha}$, that is, condition (7) for $c=\frac{(d-1)q}{2}$ holds.
Now the proof of $(b)$ follows from Theorem 1 $(a)$ and Theorem 2$(a),$ since
either $\gamma+t$ $\in U(\rho^{\alpha_{1}},p)$ or $\gamma+t\in$ $E_{k}%
\backslash E_{k+1}$ for $k=1,2,...,d-1$ ( see (42) in Remark 1)
\end{proof}

\begin{remark}
Here we note that the non-resonance domain $U(c_{8}\rho^{\alpha_{1}},p)$ has
an asymptotically full measure on $\mathbb{R}^{d}$ in the sence that
$\frac{\mu(U\cap B(\rho))}{\mu(B(\rho))}$ tends to $1$ as $\rho$ tends to
infinity, where $B(\rho)=\{x\in\mathbb{R}^{d}:\mid x\mid=\rho\}.$ Clearly,
$B(\rho)\cap V_{b}(c_{8}\rho^{\alpha_{1}})$ is the part of sphere $B(\rho),$
which is contained between two parallel hyperplanes

$\{x:\mid x\mid^{2}-\mid x+b\mid^{2}=-c_{8}\rho^{\alpha_{1}}\}$ and $\{x:\mid
x\mid^{2}-\mid x+b\mid^{2}=c_{8}\rho^{\alpha_{1}}\}.$ The distance of these
hyperplanes from origin is $O(\frac{\rho^{\alpha_{1}}}{\mid b\mid}).$
Therefore, the relations $\mid\Gamma(p\rho^{\alpha})\mid=O(\rho^{d\alpha}),$
and $\alpha_{1}+d\alpha<1-\alpha$ ( see the first inequality in (14)) imply
\begin{align}
\mu(B(\rho)\cap V_{b}(c_{8}\rho^{\alpha_{1}}))  &  =O(\frac{\rho^{\alpha
_{1}+d-2}}{\mid b\mid}),\text{ }\mu(E_{1}\cap B(\rho))=O(\rho^{d-1-\alpha})\\
\mu(U(c_{8}\rho^{\alpha_{1}},p)\cap B(\rho))  &  =(1+O(\rho^{-\alpha}%
))\mu(B(\rho)).
\end{align}
If $x\in\cap_{i=1}^{d}V_{\gamma_{i}}(\rho^{\alpha_{d}}),$ then (31) holds for
$k=d$ and $n=1,2,...,d.$ Hence we have $\mid x\mid=O(\rho^{\alpha
_{d}+(d-1)\alpha}).$ It is imposible, because of $\alpha_{d}+(d-1)\alpha<1$ (
see the \ sixth inequality in (14)) and $x\in B(\rho).$ It means that
$(\cap_{i=1}^{d}V_{\gamma_{i}}(\rho^{\alpha_{k}}))\cap B(\rho_{0})=\emptyset$
for $\rho_{0}\gg1$. Thus, for $\rho_{0}\gg1,$ we have
\begin{equation}
\mathbb{R}^{d}\cap\{\mid x\mid>\rho_{0}\}=(U\cup(\cup_{s=1}^{d-1}%
(E_{s}\backslash E_{s+1})))\cap\{\mid x\mid>\rho_{0}\}.
\end{equation}

\end{remark}

\begin{remark}
Here we note some properties of the known parts

$\mid\gamma+t\mid^{2}+F_{k}(\gamma+t)$ (see Theorem 1)and $\lambda_{j}%
(\gamma+t)$ ( see Theorem 2) of the non-resonance and resonance eigenvalues of
$L_{t}(q(x))$. Denoting $\gamma+t$ by $x$ , where $\mid\gamma+t\mid\sim\rho,$
$\gamma+t\in U(\rho^{\alpha_{1}},p),$ we prove
\begin{equation}
\frac{\partial F_{k}(x)}{\partial x_{i}}=O(\rho^{-2\alpha_{1}+\alpha}),\forall
i=1,2,...,d;\forall k=1,2,...
\end{equation}
by induction on $k.$ For $k=1$ the formula (43) follows from (4) and
\begin{equation}
\frac{\partial}{\partial x_{i}}(\dfrac{1}{\mid x\mid^{2}-\mid x-\gamma_{1}%
\mid^{2}})=\dfrac{-2\gamma_{1}(i)}{(\mid x\mid^{2}-\mid x-\gamma_{1}\mid
^{2})^{2}}=O(\rho^{-2\alpha_{1}+\alpha}),
\end{equation}
where $\gamma_{1}(i)$ is the $i$-th component of the vector $\gamma_{1}%
\in\Gamma(p\rho^{\alpha})$ hence is equal to $O(\rho^{\alpha}).$ Now suppose
that (43) holds for $k=s.$ Using this and (23), replacing $\mid x\mid^{2}$ by
$\mid x\mid^{2}+F_{s}(x)$ in (44) and evaluting as above we obtain
\[
\frac{\partial}{\partial x_{i}}(\dfrac{1}{\mid x\mid^{2}+F_{s}-\mid
x-\gamma_{1}\mid^{2}})=\dfrac{-2\gamma_{1}(i)+\frac{\partial F_{s}%
(x)}{\partial x_{i}}}{(\mid x\mid^{2}+F_{s}-\mid x-\gamma_{1}\mid^{2})^{2}%
}=O(\rho^{-2\alpha_{1}+\alpha}).
\]
This formula together with the definition of $F_{k}$ give (43) for $k=s+1.$

Now denoting $\lambda_{i}(\gamma+t)-\mid\gamma+t\mid^{2}$ by $r_{i}(\gamma+t)$
we prove that
\begin{equation}
\mid r_{i}(x)-r_{i}(x^{^{\prime}})\mid\leq2\rho^{\frac{1}{2}\alpha_{d}}\mid
x-x^{^{\prime}}\mid,\forall i.
\end{equation}
Clearly $r_{1}(x)\leq r_{2}(x)\leq...\leq$ $r_{b_{k}}(x)$ are the eigenvalue
of the matrix

$C(x)-\mid x\mid^{2}I\equiv C^{^{\prime}}(x),$ where $C(x)$ is defined in
(26). By definition, only the diagonal elements of the matrix $C^{^{\prime}%
}(x)$ depend on $x$ and they are

$\mid x\mid^{2}-\mid x-a_{i}\mid^{2}=2(x,a_{i})-\mid a_{i}\mid^{2},$ where
$a_{i}=h_{i}+t-\gamma-t$ and $h_{i}+t\in B_{k}(\gamma+t,p_{1}).$ It follows
from the definitions of $B_{k}(\gamma+t,p_{1})$ ( for $k<d$ see (25)), and
$\alpha_{d}$ ( see introduction) that $\mid a_{i}\mid<\frac{1}{2}\rho
^{\frac{1}{2}\alpha_{k}}+p_{1}\rho^{\alpha}<\rho^{\frac{1}{2}\alpha_{d}}$.
Using this and taking into account that $C^{^{\prime}}(x)-C^{^{\prime}%
}(x^{^{\prime}})=(a_{i,j}),$ where $a_{i,i}=2(x-x^{^{\prime}},a_{i}),$
$a_{i,j}=0$ for $i\neq j,$ we obtain $\parallel C^{^{\prime}}(x)-C^{^{\prime}%
}(x^{^{\prime}})\parallel\leq2\rho^{\frac{1}{2}\alpha_{d}}\mid x-x^{^{\prime}%
}\mid$ from which follows (45).
\end{remark}

\section{Asymptotic Formulas for Bloch Functions}

In this section using the asymptotic formulas for eigenvalues and the
simplicity conditions (11), (12), we prove the asymptotic formulas for the
Bloch functions with a quasimomenta of the simple set $B$.

\begin{theorem}
If $\gamma+t\in B$ and $\mid\gamma+t\mid\sim\rho,$ then there exists a unique
eigenvalue $\Lambda_{N}(t)$ satisfying (5) for $k=1,2,...,[\frac{p}{3}],$
where $p$ is defined in (3). This is a simple eigenvalue and the corresponding
eigenfunction $\Psi_{N,t}(x)$ of $L(q(x))$ satisfies (9) if $q(x)\in
W_{2}^{s_{0}}(F),$ where $s_{0}=\frac{3d-1}{2}(3^{d}+d+2)+\frac{1}{4}%
d3^{d}+d+6.$
\end{theorem}

\begin{proof}
By Theorem 1(b) if $\gamma+t\in B\subset U(\rho^{\alpha_{1}},p),$ then there
exists an eigenvalue $\Lambda_{N}(t)$ satisfying (5) for $k=1,2,...,[\frac
{1}{3}(p-\frac{1}{2}q(d-1))].$ Since

$k_{1}=[\frac{d}{3\alpha}]+2\leq\frac{1}{3}(p-\frac{1}{2}q(d-1))$ (see the
third inequality in (14)) formula (5) holds for $k=k_{1}.$ Therefore using
(5), the relation $3k_{1}\alpha>d+2\alpha$ ( see the fifth inequality in
(14)), and notations $F(\gamma+t)=\mid\gamma+t\mid^{2}+F_{k_{1}-1}(\gamma+t)$,
$\varepsilon_{1}=\rho^{-d-2\alpha}$ ( see Step 1 in introduction), we obtain
\begin{equation}
\Lambda_{N}(t)=F(\gamma+t)+o(\varepsilon_{1}).
\end{equation}
Let $\Psi_{N}$ be any normalized eigenfunction corresponding to $\Lambda_{N}$.
Since the normalized eigenfunction is defined up to constant of modulas $1,$
without loss of generality it can assumed that $\arg b(N,\gamma)=0,$ where
$b(N,\gamma)=(\Psi_{N},e^{i(\gamma+t,x)}).$ Therefore to prove (9) it suffices
to show that (13) holds. To prove (13) first \ we estimate \ $\sum
_{\gamma^{^{\prime}}\notin K}\mid b(N,\gamma^{^{\prime}})\mid^{2}$and then
$\sum_{\gamma^{^{\prime}}\in K\backslash\{\gamma\}}\mid b(N,\gamma^{^{\prime}%
})\mid^{2},$ where $K$ is defined in (11), (12). Using (46), the definition of
$K$, and (15), we get
\begin{align}
&  \mid\Lambda_{N}-\mid\gamma^{^{\prime}}+t\mid^{2}\mid>\frac{1}{4}%
\rho^{\alpha_{1}},\text{ }\forall\gamma^{^{\prime}}\notin K,\\
\sum_{\gamma^{^{\prime}}\notin K}  &  \mid b(N,\gamma^{^{\prime}})\mid
^{2}=\parallel q(x)\Psi_{N}\parallel^{2}O(\rho^{-2\alpha_{1}})=O(\rho
^{-2\alpha_{1}}).\nonumber
\end{align}
If $\gamma^{^{\prime}}\in K$ , then by (46) and by definition of $K,$ it
follows that
\begin{equation}
\mid\Lambda_{N}-\mid\gamma^{^{\prime}}+t\mid^{2}\mid<\frac{1}{2}\rho
^{\alpha_{1}}%
\end{equation}
Now we prove that the simplicity conditions (11), (12) imply
\begin{equation}
\mid b(N,\gamma^{^{\prime}})\mid\leq c_{4}\rho^{-c\alpha},\text{ }%
\forall\gamma^{^{\prime}}\in K\backslash\{\gamma\},
\end{equation}
where $c=p-dq-\frac{1}{4}d3^{d}-3.$ If for $\gamma^{^{\prime}}+t\in
U(\rho^{\alpha_{1}},p)$ and $\gamma^{^{\prime}}\in K\backslash\{\gamma\}$ the
inequality in (49) is not true, then by (48) and Theorem 1(a), we have
\begin{equation}
\Lambda_{N}=\mid\gamma^{^{\prime}}+t\mid^{2}+F_{k-1}(\gamma^{^{\prime}%
}+t)+O(\rho^{-3k\alpha})
\end{equation}
for $k=1,2,...,[\frac{1}{3}(p-c)]=[\frac{1}{3}(dq+\frac{1}{4}d3^{d}+3)].$
Since $\alpha=\frac{1}{q}$ and

$k_{1}\equiv\lbrack\frac{d}{3\alpha}]+2<\frac{1}{3}(dq+\frac{1}{4}d3^{d}+3)$,
\ the formula (50) holds for $k=k_{1}.$ Therefore arguing as in the prove of
(46), we get $\Lambda_{N}-F(\gamma^{^{\prime}}+t)=o(\varepsilon_{1})$. \ This
with (46) contradicts (11). Similarly, if the inequality in (49) does not hold
for $\gamma^{^{\prime}}+t\in(E_{k}\backslash E_{k+1})$ and $\gamma^{^{\prime}%
}\in K,$ then by Theorem 2(a)
\begin{equation}
\Lambda_{N}=\lambda_{j}(\gamma^{^{\prime}}+t)+O(\rho^{-(p-c-\frac{1}{4}%
d3^{d})\alpha}),
\end{equation}
where $(p-c-\frac{1}{4}d3^{d})\alpha=(dq+3)\alpha>d+2\alpha$ . Hence we have

$\Lambda_{N}-\lambda_{j}(\gamma^{^{\prime}}+t)=o(\varepsilon_{1}).$ This with
(46) contradicts (12). So the inequality in (49) holds. Therefore, using $\mid
K\mid=O(\rho^{d-1}),$ $q\alpha=1,$ we get
\begin{equation}
\sum_{\gamma^{^{\prime}}\in K\backslash\{\gamma\}}\mid b(N,\gamma^{^{\prime}%
})\mid^{2}=O(\rho^{-(2c-q(d-1))\alpha})=O(\rho^{-(2p-(3d-1)q-\frac{1}{2}%
d3^{d}-6)\alpha}).
\end{equation}
If $s=s_{0},$ that is, $p=s_{0}-d$ then $2p-(3d-1)q-\frac{1}{2}d3^{d}-6=6.$
Since $\alpha_{1}=3\alpha,$ the equality (52) and the equality in (47) imply
(13). Thus we proved that the equality (9) holds for any normalized
eigenfunction $\Psi_{N}$ corresponding to any eigenvalue $\Lambda_{N}$
\ satisfying (5). If there exist two different eigenvalues or multiple
eigenvalue satisfying (5), then there exist two orthogonal normalized
eigenfunction satisfying (9), which is imposible. Therefore $\Lambda_{N}$\ is
a simple eigenvalue. It follows from Theorem 1(a) that $\Lambda_{N}$ satisfies
(5) for $k=1,2,...,[\frac{p}{3}],$ because the inequality (7) holds for $c=0$
( see (9)).
\end{proof}

\begin{remark}
Since for $\gamma+t\in B$ \ there exists a unique eigenvalue satisfying (5),
(46) we denote this eigenvalue by $\Lambda(\gamma+t).$ Since this eigenvalue
is simple, we denote the \ corresponding eigenfunction by $\Psi_{\gamma
+t}(x).$ By Theorem 3 this eigenfunction satisfies (9). Clearly, for
$\gamma+t\in B$ \ there exists a unique index $N\equiv N(\gamma+t)$ such that
$\Lambda(\gamma+t)=\Lambda_{N(\gamma+t)}$) and $\Psi_{\gamma+t}(x)=\Psi
_{N(\gamma+t)}(x)).$
\end{remark}

Now we prove the asymptotic formulas of arbitrary order for $\Psi_{\gamma
+t}(x).$

\begin{theorem}
If $\gamma+t\in B$ and $\mid\gamma+t\mid\sim\rho,$ then the eigenfunction

$\Psi_{\gamma+t}(x)\equiv\Psi_{N(\gamma+t)}(x)$ corresponding to the
eigenvalue $\Lambda_{N}\equiv\Lambda(\gamma+t)$ satisfies formulas (10), for
$k=1,2,...,n$, where $n=[\frac{1}{6}(2p-(3d-1)q-\frac{1}{2}d3^{d}-6)],$
$\Phi_{0}(x)=0,$ $\Phi_{1}(x)=%
{\displaystyle\sum_{\gamma_{1}\in\Gamma(\rho^{\alpha})}}
\dfrac{q_{\gamma_{1}}e^{i(\gamma+t+\gamma_{1},x)}}{(\mid\gamma+t\mid^{2}%
-\mid\gamma+\gamma_{1}+t\mid^{2})},$

and $\Phi_{k-1}(x)$ for $k>2$ is a linear combination of $e^{i(\gamma
+t+\gamma^{^{\prime}},x)}$ for

$\gamma^{^{\prime}}\in\Gamma((k-1)\rho^{\alpha})\cup\{0\}$ with coefficients
(58), (59).
\end{theorem}

\begin{proof}
By Theorem 3, formula (10) for $k=1$ is proved. To prove formula (10) for
arbitrary $k\leq n$ we prove the following equivalent relations
\begin{equation}
\sum_{\gamma^{^{\prime}}\in\Gamma^{c}(k-1)}\mid b(N,\gamma+\gamma^{^{\prime}%
})\mid^{2}=O(\rho^{-2k\alpha_{1}}),
\end{equation}%
\begin{equation}
\Psi_{N}=b(N,\gamma)e^{i(\gamma+t,x)}+\sum_{\gamma^{^{\prime}}\in
\Gamma((k-1)\rho^{\alpha})}b(N,\gamma+\gamma^{^{\prime}})e^{i(\gamma
+t+\gamma^{^{\prime}},x)}+H_{k}(x),
\end{equation}
where $\Gamma^{c}(m)\equiv\Gamma\backslash(\Gamma(m\rho^{\alpha})\cup\{0\})$
and $\parallel H_{k}(x)\parallel=O(\rho^{-k\alpha_{1}}).$ The case $k=1$ is
proved due to (13). Assume that (53) is true for $k=m$ . Then using (54) for
$\ \ \ k=m,$ and (3), we have $\Psi_{N}(x)(q(x))=H(x)+O(\rho^{-m\alpha_{1}}),$
where $H(x)$ is a linear combination of $e^{i(\gamma+t+\gamma^{^{\prime}},x)}$
for $\gamma^{^{\prime}}\in\Gamma(m\rho^{\alpha})\cup\{0\}.$ Hence
$(H(x),e^{i(\gamma+t+\gamma^{^{\prime}},x)})=0$ for $\gamma^{^{\prime}}%
\in\Gamma^{c}(m).$ So using (15) and the inequality in (47), we get%
\begin{equation}
\sum_{\gamma^{^{\prime}}}\mid b(N,\gamma+\gamma^{^{\prime}})\mid^{2}%
=\sum_{\gamma^{^{\prime}}}\mid\dfrac{(O(\rho^{-m\alpha_{1}}),e^{i(\gamma
+t+\gamma^{^{\prime}},x)})}{\Lambda_{N}-\mid\gamma+\gamma^{^{\prime}}%
+t\mid^{2}}\mid^{2}=O(\rho^{-2(m+1)\alpha_{1}}),
\end{equation}
where the sum is taken under conditions $\gamma^{^{\prime}}\in\Gamma^{c}(m)$,
$\gamma+\gamma^{^{\prime}}\notin K.$ On the other hand, using $\alpha
_{1}=3\alpha,$ (52), and the definition of $n$ ( see Theorem 4), we get
\[
\sum_{\gamma^{^{\prime}}\in K\backslash\{\gamma\}}\mid b(N,\gamma^{^{\prime}%
})\mid^{2}=O(\rho^{-2n\alpha_{1}}).
\]
This with (55) implies (53) for $k=m+1.$ Thus (54) is also proved. Here
$b(N,\gamma)$ and $b(N,\gamma+\gamma^{^{\prime}})$ for $\gamma^{^{\prime}}%
\in\Gamma((n-1)\rho^{\alpha})$ can be calculated as follows. First we express
$b(N,\gamma+\gamma^{^{\prime}})$ by $b(N,\gamma)$. For this we apply (17) for
$b(N,\gamma+\gamma^{^{\prime}}),$ where $\gamma^{^{\prime}}\in\Gamma
((n-1)\rho^{\alpha}),$ that is, in (17) replace $\gamma^{^{\prime}}$ by
$\gamma+\gamma^{^{\prime}}$. Iterate it $n$ times and every times isolate the
terms with multiplicand $b(N,\gamma).$ In other word apply (17) for
$b(N,\gamma+\gamma^{^{\prime}})$ and isolate the terms with multiplicand
$b(N,\gamma).$ Then apply (17) for $b(N,\gamma+\gamma^{^{\prime}}-\gamma_{1})$
when $\gamma^{^{\prime}}-\gamma_{1}\neq0.$ Then apply (17) for

$b(N,\gamma+\gamma^{^{\prime}}-\sum_{i=1}^{2}\gamma_{i})$ when $\gamma
^{^{\prime}}-\sum_{i=1}^{2}\gamma_{i}\neq0,$ etc. Apply (17) for

$b(N,\gamma+\gamma^{^{\prime}}-\sum_{i=1}^{j}\gamma_{i})$ when $\gamma
^{^{\prime}}-\sum_{i=1}^{j}\gamma_{i}\neq0,$ where $\gamma_{i}\in\Gamma
(\rho^{\alpha}),$

$j=3,4,...,n-1.$ Then using (4) and the relations

$\mid\Lambda_{N}-\mid\gamma+t+\gamma^{^{\prime}}-\sum_{i=1}^{j}\gamma_{i}%
\mid^{2}\mid>\frac{1}{2}\rho^{\alpha_{1}}$ ( see (19) and take into account that

$\gamma^{^{\prime}}-\sum_{i=1}^{j}\gamma_{i}\in\Gamma(p\rho^{\alpha}),$ since
$p>2n$), $\Lambda_{N}=P(\gamma+t)+O(\rho^{-n\alpha_{1}}),$ where
$P(\gamma+t)=\mid\gamma+t\mid^{2}+F_{[\frac{p}{3}]}(\gamma+t)$ ( see Theorem
3), we obtain%
\begin{equation}
b(N,\gamma+\gamma^{^{\prime}})=\sum_{k=1}^{n-1}A_{k}(\gamma^{^{\prime}%
})b(N,\gamma)+O(\rho^{-n\alpha_{1}}),
\end{equation}
where

$A_{1}(\gamma^{^{\prime}})=\dfrac{q_{\gamma^{^{\prime}}}}{P(\gamma
+t)-\mid\gamma+\gamma^{^{\prime}}+t\mid^{2}}=\dfrac{q_{\gamma^{^{\prime}}}%
}{\mid\gamma+t\mid^{2}-\mid\gamma+\gamma^{^{\prime}}+t\mid^{2}}+O(\frac
{1}{\rho^{3\alpha_{1}}}),$%

\[
A_{k}(\gamma^{^{\prime}})=%
{\displaystyle\sum_{\gamma_{1},...,\gamma_{k-1}}}
\dfrac{q_{\gamma_{1}}q_{\gamma_{2}}...q_{\gamma_{k-1}}q_{\gamma^{^{\prime}%
}-\gamma_{1}-\gamma_{2}-...-\gamma_{k-1}}}{\prod_{j=0}^{k-1}(P(\gamma
+t)-\mid\gamma+t+\gamma^{^{\prime}}-\sum_{i=1}^{j}\gamma_{i}\mid^{2})}%
=O(\rho^{-k\alpha_{1}}),
\]%
\begin{equation}
\sum_{\gamma^{\ast}\in\Gamma((n-1)\rho^{\alpha})}\mid A_{1}(\gamma^{\ast}%
)\mid^{2}=O(\rho^{-2\alpha_{1}}),\sum_{\gamma^{\ast}\in\Gamma((n-1)\rho
^{\alpha})}\mid A_{k}(\gamma^{\ast})\mid=O(\rho^{-k\alpha_{1}})
\end{equation}
for $k>1.$ Now from (54) for $k=n$ and (56), we obtain
\begin{align*}
\Psi_{N}(x)  &  =b(N,\gamma)e^{i(\gamma+t,x)}+\\
&  \sum_{\gamma^{\ast}\in\Gamma((n-1)\rho^{\alpha})}\sum_{k=1}^{n-1}%
(A_{k}(\gamma^{\ast})b(N,\gamma)+O(\rho^{-n\alpha_{1}}))e^{i(\gamma
+t+\gamma^{\ast},x)})+H_{n}(x).
\end{align*}
Using the equalities $\parallel\Psi_{N}\parallel=1,$ $\arg b(N,\gamma)=0,$
$\parallel H_{n}\parallel=O(\rho^{-n\alpha_{1}})$ and taking into account that
the functions $e^{i(\gamma+t,x)},$ $H_{n}(x),$ $e^{i(\gamma+t+\gamma^{\ast
},x)},$ $(\gamma^{\ast}\in\Gamma((n-1)\rho^{\alpha}))$ are orthogonal, we get

$1=\mid b(N,\gamma)\mid^{2}+\sum_{k=1}^{n-1}(\sum_{\gamma^{\ast}\in
\Gamma((n-1)\rho^{\alpha})}\mid A_{k}(\gamma^{\ast})b(N,\gamma)\mid^{2}%
+O(\rho^{-n\alpha_{1}})),$
\begin{equation}
b(N,\gamma)=(1+\sum_{k=1}^{n-1}(\sum_{\gamma^{\ast}\in\Gamma((n-1)\rho
^{\alpha})}\mid A_{k}(\gamma^{\ast})\mid^{2}))^{-\frac{1}{2}}+O(\rho
^{-n\alpha_{1}}))
\end{equation}
(see the second equality in (57)). Thus from (56), we obtain
\begin{equation}
b(N,\gamma+\gamma^{^{\prime}})=(\sum_{k=1}^{n-1}A_{k}(\gamma^{^{\prime}%
}))(1+\sum_{k=1}^{n-1}\sum_{\gamma^{\ast}}\mid A_{k}(\gamma^{\ast})\mid
^{2})^{-\frac{1}{2}}+O(\rho^{-n\alpha_{1}}).
\end{equation}
Consider the case $n=2.$ By (58), (57), (59) we have $b(N,\gamma
)=1+O(\rho^{-2\alpha_{1}}),$

$b(N,\gamma+\gamma^{^{\prime}})=A_{1}(\gamma^{^{\prime}})+O(\rho^{-2\alpha
_{1}})=\dfrac{q_{\gamma^{^{\prime}}}}{\mid\gamma+t\mid^{2}-\mid\gamma
+\gamma^{^{\prime}}+t\mid^{2}}+O(\rho^{-2\alpha_{1}})$ for all $\gamma
^{^{\prime}}\in\Gamma(\rho^{\alpha}).$ These and (54) for $k=2$ imply the
formula for $\Phi_{1}$
\end{proof}

\section{Simple Sets and Fermi Surfaces}

In this section we consider the simple sets $B$ and construct a big part of
the isoenergetic surfaces corresponding to $\rho^{2}$ for big $\rho.$ The
isoenergetic surfaces of $L(q)$ corresponding to $\rho^{2}$ is the set
$I_{\rho}(q(x))=\{t\in F^{\ast}:\exists N,\Lambda_{N}(t)=\rho^{2}\}.$ In the
case $q(x)=0$ \ the isoenergetic surface $I_{\rho}(0)=$ $\{t\in F^{\ast
}:\exists\gamma\in\Gamma,\mid\gamma+t\mid^{2}=\rho^{2}\}$ is the translation
of the sphere

$B(\rho)=\{\gamma+t:t\in F^{\ast},\gamma\in\Gamma,\mid\gamma+t\mid^{2}%
=\rho^{2}\}$ by the vectors $\gamma\in\Gamma.$ We call $B(\rho)$ the
translated isoenergetic surfaces of $L(0)$ corresponding to $\rho^{2}.$
Similarly, we call the sets $P_{\rho}^{^{\prime}}=\{\gamma+t:\Lambda
(\gamma+t)=\rho^{2}\}$ and

$P_{\rho}^{^{\prime\prime}}=\{t\in F^{\ast}:\exists\gamma\in\Gamma
,\Lambda(\gamma+t)=\rho^{2}\},$ where $\Lambda(\gamma+t)=\Lambda_{N(\gamma
+t)}(t)$ is defined in Remark 3, the parts of translated isoenergetic surfaces
and isoenergetic surfaces of $L(q).$ In this section we construct the subsets
$I_{\rho}^{^{\prime}}$ and $I_{\rho}^{^{\prime\prime}}$ of $P_{\rho}%
^{^{\prime}}$ and $P_{\rho}^{^{\prime\prime}}$ respectively and prove that the
measures of these subsets are asymptotically equal to the measure of the
isoenergetic surfaces $I_{\rho}(0)$ of $L(0)$. In other word we construct a
big part (in some sense) of isoenergetic surfaces $I_{\rho}(q(x))$ of $L(q)$.
As we see below the set $I_{\rho}^{^{\prime\prime}}$ is a translation of
$I_{\rho}^{^{\prime}}$ by vectors $\gamma\in\Gamma$ to $F^{\ast}$ and the set
$I_{\rho}^{^{\prime}}$ lies in $\varepsilon$ nieghborhood of the surface
$S_{\rho}=\{x\in U(2\rho^{\alpha_{1}},p):F(x)=\rho^{2}\},$ where $F(x)$ is
defined in Step 1 of introduction. Due to (46) ( replace $\gamma+t$ by $x$ in
$F(\gamma+t)$ ) it is natural to call $S_{\rho}$ the approximated isoenergetic
surfaces in the nonresonance domain. Here we construct a part of the simple
set $B$ in nieghborhood of $S_{\rho}$ that containes $I_{\rho}^{^{\prime}}$.
For this we consider the surface $S_{\rho}$. As we noted in introduction \ (
see Step 2 and (11)) the non-resonance eigenvalue $\Lambda(\gamma+t)$ does not
coincide with other non-resonance eigenvalue $\Lambda(\gamma+t+b)$ if $\mid
F(\gamma+t)-F(\gamma+t+b)\mid>2\varepsilon_{1}$ for $\gamma+t+b\in
U(\rho^{\alpha_{1}},p)$ and $b\in\Gamma\backslash\{0\}$. Therefore we
eliminate
\begin{equation}
P_{b}=\{x:x,x+b\in U(\rho^{\alpha_{1}},p),\mid F(x)-F(x+b)\mid<3\varepsilon
_{1}\}
\end{equation}
for $b\in\Gamma\backslash\{0\}$ from $S_{\rho}$. Denote the remaining part of
$S_{\rho}$ by $S_{\rho}^{^{\prime}}.$ Then we consider the $\varepsilon$
neighbourhood $U_{\varepsilon}(S_{\rho}^{^{\prime}})=\cup_{a\in S_{\rho
}^{^{\prime}}}U_{\varepsilon}(a)\}$ of $S_{\rho}^{^{\prime}}$ ,

where $\varepsilon=\frac{\varepsilon_{1}}{7\rho},$ $U_{\varepsilon}(a)=\{x\in
R^{d}:\mid x-a\mid<\varepsilon\}.$ In this set the first simplicity condition
(11) holds (see Lemma 2(a)). Denote by

$Tr(E)=\{\gamma+x\in U_{\varepsilon}(S_{\rho}^{^{\prime}}):\gamma\in
\Gamma,x\in E\}$ and

$Tr_{F^{\star}}(E)\equiv\{\gamma+x\in F^{\star}:\gamma\in\Gamma,x\in E\}$ the
translations of $E\subset R^{d}$ into $U_{\varepsilon}(S_{\rho}^{^{\prime}})$
and $F^{\star}$ respectively. In order that the second simplicity condition
(12) holds, we discart from $U_{\varepsilon}(S_{\rho}^{^{\prime}})$ the
translation $Tr(A(\rho))$ of
\begin{equation}
A(\rho)\equiv\cup_{k=1}^{d-1}(\cup_{\gamma_{1},\gamma_{2},...,\gamma_{k}%
\in\Gamma(p\rho^{\alpha})}(\cup_{i=1}^{b_{k}}A_{k,i}(\gamma_{1},\gamma
_{2},...,\gamma_{k}))),
\end{equation}

where $A_{k,i}(\gamma_{1},...,\gamma_{k})=$

$\{x\in(\cap_{i=1}^{k}V_{\gamma_{i}}(\rho^{\alpha_{k}})\backslash E_{k+1})\cap
K_{\rho}:\lambda_{i}(x)\in(\rho^{2}-3\varepsilon_{1},\rho^{2}+3\varepsilon
_{1})\},$

$\lambda_{i}(x),$ $b_{k}$ is defined in Theorem 2 and
\begin{equation}
K_{\rho}=\{x\in\mathbb{R}^{d}:\mid\mid x\mid^{2}-\rho^{2}\mid<\rho^{\alpha
_{1}}\}.
\end{equation}
As a result we construct the part $U_{\varepsilon}(S_{\rho}^{^{\prime}%
})\backslash Tr(A(\rho))$ of the simple set $B$ (see Theorem 5(a)) which
contains the set $I_{\rho}^{^{\prime}}$ (see Theorem 5(c)). For this we need
the following lemma.

\begin{lemma}
$(a)$ If $x\in U_{\varepsilon}(S_{\rho}^{^{\prime}})$ and $x+b\in
U(\rho^{\alpha_{1}},p),$ where $b\in\Gamma,$ then
\begin{equation}
\mid F(x)-F(x+b)\mid>2\varepsilon_{1},
\end{equation}
where $\varepsilon=\frac{\varepsilon_{1}}{7\rho},\varepsilon_{1}%
=\rho^{-d-2\alpha},$ $F(x)=\mid x\mid^{2}+F_{k_{1}-1}(x),$ $k_{1}=[\frac
{d}{3\alpha}]+2,$ hence for $\gamma+t\in U_{\varepsilon}(S_{\rho}^{^{\prime}%
})$ the simplisity condition (11) holds.

$(b)$ If $x\in U_{\varepsilon}(S_{\rho}^{^{\prime}}),$ then $x+b\notin
U_{\varepsilon}(S_{\rho}^{^{\prime}})$ for all $b\in\Gamma$ $.$

$(c)$If $E$ is a bounded subset of $\mathbb{R}^{d}$, then $\mu(Tr(E))\leq
\mu(E)$.

$(d)$ If $E\subset U_{\varepsilon}(S_{\rho}^{^{\prime}}),$ then $\mu
(Tr_{F^{\star}}(E))=\mu(E).$
\end{lemma}

\begin{proof}
$(a)$ If $x\in U_{\varepsilon}(S_{\rho}^{^{\prime}}),$ then there exists a
point $a$ in $S_{\rho}^{^{\prime}}$ such that $x\in U_{\varepsilon}(a)$. Since
$S_{\rho}^{^{\prime}}\cap P_{b}=\emptyset$ ( see (60) and the definition of
$S_{\rho}^{^{\prime}}$), we have
\begin{equation}
\mid F(a)-F(a+b)\mid\geq3\varepsilon_{1}%
\end{equation}
On the other hand, using (43) and the obvious relations

$\mid x\mid<\rho+1,$ $\mid x-a\mid<\varepsilon,$ $\mid x+b-a-b\mid
<\varepsilon,$ we obtain
\begin{equation}
\mid F(x)-F(a)\mid<3\rho\varepsilon,\mid F(x+b)-F(a+b)\mid<3\rho\varepsilon
\end{equation}
These inequalities together with (64) give (63), since $6\rho\varepsilon
<\varepsilon_{1}.$

$(b)$ If $x$ and $x+b$ lie in $U_{\varepsilon}(S_{\rho}^{^{\prime}}),$ then
there exist points $a$ and $c$ in $S_{\rho}^{^{\prime}}$ such that $x\in
U_{\varepsilon}(a)$ and $x+b\in U_{\varepsilon}(c).$ Repeating the proof of
(65), we get

$\mid F(c)-F(x+b)\mid<3\rho\varepsilon.$ This, the first inequality in (65),
and the relations $F(a)=\rho^{2},$ $F(c)=\rho^{2}$ (see the definition of
$S_{\rho})$ give

$\mid F(x)-F(x+b)\mid<\varepsilon_{1},$ which contradicts (63).

$(c)$ Clearly, for any bounded set $E$ there exist only a finite number of
vectors $\gamma_{1},\gamma_{2},...,\gamma_{s}$ such that $E(k)\equiv
(E+\gamma_{k})\cap U_{\varepsilon}(S_{\rho}^{^{\prime}})\neq\emptyset$ for
$k=1,2,...,s$ and $Tr(E)$ is the union of the sets $E(k)$. For $E(k)-\gamma
_{k}$ we have the relations $\mu(E(k)-\gamma_{k})=\mu(E(k)),$ $E(k)-\gamma
_{k}\subset E.$ Moreover, by $(b)$

$(E(k)-\gamma_{k})\cap(E(j)-\gamma_{j})=\emptyset$ for $k\neq j.$ Therefore
$(c)$ is true.

$(d)$ Now let $E\subset U_{\varepsilon}(S_{\rho}^{^{\prime}}).$ Then by $(b)$
the set $E$ can be devided into a finite number of the pairwise disjoint sets
$E_{1},E_{2},...,E_{n}$ such that there exist the vectors $\gamma_{1}%
,\gamma_{2},...,\gamma_{n}$ satisfying $(E_{k}+\gamma_{k})\subset F^{\star},$
$(E_{k}+\gamma_{k})\cap(E_{j}+\gamma_{j})\neq\emptyset$ for $k,j=1,2,...,n$
and $k\neq j.$ Using $\mu(E_{k}+\gamma_{k})=\mu(E_{k}),$ we get the proof of
$(d),$ because $Tr_{F^{\star}}(E)$ and $E$ are union of the pairwise disjoint
sets $E_{k}+\gamma_{k}$ and $E_{k}$ for $k=1,2,...,n$ respectively
\end{proof}

\begin{theorem}
$(a)$ The set $U_{\varepsilon}(S_{\rho}^{^{\prime}})\backslash Tr(A(\rho))$ is
a subset of $B$. For every connected open subset $E$ of $U_{\varepsilon
}(S_{\rho}^{^{\prime}})\backslash Tr(A(\rho)$ there exists a unique index $N$
such that $\Lambda_{N}(t)=\Lambda(\gamma+t)$ for $\gamma+t\in E,$ where
$\Lambda(\gamma+t)$ is defined in Remark 3. Moreover,
\begin{equation}
\frac{\partial}{\partial t_{j}}\Lambda(\gamma+t)=\frac{\partial}{\partial
t_{j}}\mid\gamma+t\mid^{2}+O(\rho^{1-2\alpha_{1}}),\forall j=1,2,...,d.
\end{equation}

$(b)$ For the part $V_{\rho}\equiv S_{\rho}^{^{\prime}}\backslash
U_{\varepsilon}(Tr(A(\rho)))$ of the approximated isoenergetic surface
$S_{\rho}$ the following holds
\begin{equation}
\mu(V_{\rho})>(1-c_{9}\rho^{-\alpha}))\mu(B(\rho)).
\end{equation}
Moreover, $U_{\varepsilon}(V_{\rho})$ lies in the subset $U_{\varepsilon
}(S_{\rho}^{^{\prime}})\backslash Tr(A(\rho))$ of the simple set $B.$

$(c)$ The isoenergetic surface $I(\rho)$ contains the set $I_{\rho}%
^{^{\prime\prime}},$ which consists of the smooth surfaces and has the
measure
\begin{equation}
\mu(I_{\rho}^{^{\prime\prime}})=\mu(I_{\rho}^{^{\prime}})>(1-c_{10}%
\rho^{-\alpha})\mu(B(\rho)),
\end{equation}
where $I_{\rho}^{^{\prime}}$ is a part of the translated isoenergetic surfaces
of $L(q),$ which is contained in the subset $U_{\varepsilon}(S_{\rho
}^{^{\prime}})\backslash Tr(A(\rho))$ of the simple set $B.$ In particular the
number $\rho^{2}$ for $\rho\gg1$ lies in the spectrum of $L(q),$ that is, the
number of the gaps in the spectrum of $L(q)$ is finite, where $q(x)\in
W_{2}^{s_{0}}(\mathbb{R}^{d}/\Omega),$ $d\geq2,$ $s_{0}=\frac{3d-1}{2}%
(3^{d}+d+2)+\frac{1}{4}d3^{d}+d+6,$ and $\Omega$ is an arbitrary lattice.
\end{theorem}

\begin{proof}
$(a)$ To prove that $U_{\varepsilon}(S_{\rho}^{^{\prime}})\backslash
Tr(A(\rho))\subset B$ we need to show that for each point $\gamma+t$ of
$U_{\varepsilon}(S_{\rho}^{^{\prime}})\backslash Tr(A(\rho))$ the simplicity
conditions (11), (12) hold and $U_{\varepsilon}(S_{\rho}^{^{\prime}%
})\backslash Tr(A(\rho))\subset U(\rho^{\alpha_{1}},p).$ By lemma 2(a), the
condition (11) holds. Now we prove that (12) holds too. Since $\gamma+t\in
U_{\varepsilon}(S_{\rho}^{^{\prime}}),$ there exists $a\in S_{\rho}^{^{\prime
}}$ such that $\gamma+t\in U_{\varepsilon}(a).$ The inequality (65) and
equality $F(a)=\rho^{2}$ imply
\begin{equation}
F(\gamma+t)\in(\rho^{2}-\varepsilon_{1},\rho^{2}+\varepsilon_{1})
\end{equation}
for $\gamma+t\in U_{\varepsilon}(S_{\rho}^{^{\prime}}).$ On the other hand
$\gamma+t\notin Tr(A(\rho)).$ It means that for any $\gamma^{^{\prime}}%
\in\Gamma,$ we have $\gamma^{^{\prime}}+t\notin A(\rho).$ If $\gamma
^{^{\prime}}\in K$ and $\gamma^{^{\prime}}+t\in E_{k}\backslash E_{k+1},$ then
by definition of $K$ ( see introduction) the inequality $\mid F(\gamma
+t)-\mid\gamma^{^{\prime}}+t\mid^{2}\mid<\frac{1}{3}\rho^{\alpha_{1}}$ holds.
This and (69) imply that $\gamma^{^{\prime}}+t\in(E_{k}\backslash E_{k+1})\cap
K_{\rho}$ ( see (62) for the definition of $K_{\rho}$). Since $\gamma
^{^{\prime}}+t\notin A(\rho),$ we have $\lambda_{i}(\gamma^{^{\prime}%
}+t)\notin(\rho^{2}-3\varepsilon_{1},\rho^{2}+3\varepsilon_{1})$ for
$\gamma^{^{\prime}}\in K$ and $\gamma^{^{\prime}}+t\in E_{k}\backslash
E_{k+1}.$ Therefore (12) follows from (69). Moreover, it is clear that the
inclusion $S_{\rho}^{^{\prime}}\subset U(2\rho^{\alpha_{1}},p)$ ( see
definition of $S_{\rho}$ and $S_{\rho}^{^{\prime}}$) implies that
$U_{\varepsilon}(S_{\rho}^{^{\prime}})\subset U(\rho^{\alpha_{1}},p).$ Thus
$U_{\varepsilon}(S_{\rho}^{^{\prime}})\backslash Tr(A(\rho))\subset B.$

Now let $E$ be a connected open subset of $U_{\varepsilon}(S_{\rho}^{^{\prime
}})\backslash Tr(A(\rho)\subset B.$ By Theorem 3 and Remark 3 for $a\in
E\subset U_{\varepsilon}(S_{\rho}^{^{\prime}})\backslash Tr(A(\rho)$ there
exists a unique index $N(a)$ such that $\Lambda(a)=\Lambda_{N(a)}(a)$,
$\Psi_{a}(x)=\Psi_{N(a),a}(x)$, $\mid(\Psi_{N(a),a}(x),e^{i(a,x)})\mid
^{2}>\frac{1}{2}$ and $\Lambda(a)$ is a simple eigenvalue. On the other hand,
for fixed $N$ the functions $\Lambda_{N}(t)$ and $(\Psi_{N,t}(x),e^{i(t,x)})$
are continuous in neighborhood of $a$ if $\Lambda_{N}(a)$ is a simple
eigenvalue. Therefore for each $a\in E$ there exists a neighborhood
$U(a)\subset E$ of $a$ such that $\mid(\Psi_{N(a),y}(x),e^{i(y,x)})\mid
^{2}>\frac{1}{2}$, for $y\in U(a).$ Since for $y\in E$ there is a unique
integer $N(y)$ satisfying $\mid(\Psi_{N(y),y}(x),e^{i(y,x)})\mid^{2}>\frac
{1}{2},$ we have $N(y)=N(a)$ for $y\in U(a).$ Hence we proved that%
\begin{equation}
\forall a\in E,\exists U(a)\subset E:N(y)=N(a),\forall y\in U(a).
\end{equation}

Now let $a_{1}$ and $a_{2}$ be two points of $E$ , and let $C\subset E$ be the
arc that joins these points. (Note that the open connected subset of
$\mathbb{R}^{d}$ is arcwise connected). Let $U(y_{1}),U(y_{2}),...,U(y_{k})$
be a finite subcover of the open cover $\cup_{a\in C}U(a)$ of the compact $C,$
where $U(a)$ is a neighborhood of $a$ satisfying (70). By (70), we have
$N(y)=N(y_{i})=N_{i}$ for $y\in U(y_{i}).$ Clearly, if $U(y_{i})\cap
U(y_{j})\neq\emptyset,$ then $N_{i}=N(z)=N_{j},$ where $z\in U(y_{i})\cap
U(y_{j})$. Thus $N_{1}=N_{2}=...=N_{k}$ and $N(a_{1})=N(a_{2}).$

To calculate the partial derivatives of the function $\Lambda(\gamma
+t)=\Lambda_{N}(t)$ we write the operator $L_{t}$ in the form $-\triangle
-(2it,\nabla)+(t,t).$ Then, it is clear that
\begin{align}
\frac{\partial}{\partial t_{j}}\Lambda_{N}(t)  &  =2t_{j}(\Phi_{N,t}%
(x),\Phi_{N,t}(x))-2i(\frac{\partial}{\partial x_{j}}\Phi_{N,t}(x),\Phi
_{N,t}(x)),\\
\Phi_{N,t}(x)  &  =\sum_{\gamma^{^{\prime}}\in\Gamma}b(N,\gamma^{^{\prime}%
})e^{i(\gamma^{^{\prime}},x)},
\end{align}
where $\Phi_{N,t}(x)=e^{-i(t,x)}\Psi_{N,t}(x).$ If $\mid\gamma^{^{\prime}}%
\mid\geq2\rho,$ then using

$\Lambda_{N}\equiv\Lambda(\gamma+t)=\rho^{2}+O(\rho^{-\alpha}),$ ( see (46),
(69)), and the obvious inequality

$\mid\Lambda_{N}-\mid\gamma^{^{\prime}}-\gamma_{1}-\gamma_{2}-...-\gamma
_{k}+t\mid^{2}\mid>c_{11}\mid\gamma^{^{\prime}}\mid^{2}$ for $k=0,1,...,p,$
where $\mid\gamma_{1}\mid<\frac{1}{4p}\mid\gamma^{^{\prime}}\mid,$ and
itarating (17) $p$ times by using the decomposition $q(x)=\sum_{\mid\gamma
_{1}\mid<\frac{1}{4p}\mid\gamma^{^{\prime}}\mid}q_{\gamma_{1}}e^{i(\gamma
_{1},x)}+O(\mid\gamma^{^{\prime}}\mid^{-p}),$ we get
\begin{align}
b(N,\gamma^{^{\prime}})  &  =\sum_{\gamma_{1},\gamma_{2},...}\dfrac
{q_{\gamma_{1}}q_{\gamma_{2}}...q_{\gamma_{p}}b(N,\gamma^{^{\prime}}%
-\sum_{i=1}^{p}\gamma_{i})}{\prod_{j=0}^{p-1}(\Lambda_{N}-\mid\gamma
^{^{\prime}}-\sum_{i=1}^{j}\gamma_{i}+t\mid^{2})}+O(\mid\gamma^{^{\prime}}%
\mid^{-p}),\\
b(N,\gamma^{^{\prime}})  &  =O(\mid\gamma^{^{\prime}}\mid^{-p}),\text{
}\forall\mid\gamma^{^{\prime}}\mid\geq2\rho
\end{align}
By (74) the series in (72) can be differentiated term by term. Hence
\begin{equation}
-i(\frac{\partial}{\partial x_{j}}\Phi_{N,t},\Phi_{N,t})=\sum_{\gamma
^{^{\prime}}\in\Gamma}\gamma^{^{\prime}}(j)\mid b(N,\gamma^{^{\prime}}%
)\mid^{2}=\gamma(j)\mid b(N,\gamma)\mid^{2}+\Sigma_{1}+\Sigma_{2},
\end{equation}
where $\Sigma_{1}=\sum_{\mid\gamma^{^{\prime}}\mid\geq2\rho}\gamma^{^{\prime}%
}(j)\mid b(N,\gamma^{^{\prime}})\mid^{2},$ $\Sigma_{2}=\sum_{\mid
\gamma^{^{\prime}}\mid<2\rho,\gamma^{^{\prime}}\neq\gamma}\gamma^{^{\prime}%
}(j)\mid b(N,\gamma^{^{\prime}})\mid^{2}.$

By (13), $\sum_{2}=O(\rho^{-2\alpha_{1}+1}),$ $\gamma(j)\mid b(N,\gamma
)\mid^{2}=\gamma(j)(1+O(\rho^{-2\alpha_{1}}),$ and by (74), $\sum_{1}%
=O(\rho^{-2\alpha_{1}}).$ Therefore (71) and (75) imply (66).

$(b)$ To ptove the inclusion $U_{\varepsilon}(V_{\rho})\subset$
$U_{\varepsilon}(S_{\rho}^{^{\prime}})\backslash Tr(A(\rho))$ we need to show
that if $a\in V_{\rho},$ then $U_{\varepsilon}(a)\subset U_{\varepsilon
}(S_{\rho}^{^{\prime}})\backslash Tr(A(\rho)).$ This is clear, because the
relation $a\in V_{\rho}\subset S_{\rho}^{^{\prime}}$ implies that
$U_{\varepsilon}(a)\subset U_{\varepsilon}(S_{\rho}^{^{\prime}})$ and the
relation $a\notin U_{\varepsilon}(Tr(A(\rho)))$ implies that $U_{\varepsilon
}(a)\cap Tr(A(\rho))=\emptyset.$ To prove (67) first we estimate the measure
of $S_{\rho},S_{\rho}^{^{\prime}},U_{2\varepsilon}(A(\rho))$, namely we prove
\begin{align}
\mu(S_{\rho})  &  >(1-c_{12}\rho^{-\alpha})\mu(B(\rho)),\\
\mu(S_{\rho}^{^{\prime}})  &  >(1-c_{13}\rho^{-\alpha})\mu(B(\rho)),\\
\mu(U_{2\varepsilon}(A(\rho)))  &  =O(\rho^{-\alpha})\mu(B(\rho))\varepsilon
\end{align}
( see below, Estimations 1, 2, 3). The estimation (67) of the measure of the
set $V_{\rho}$ is done in Estimation 4 by using Estimations 1, 2, 3.

$(c)$ In Estimation 5 we prove the formula (68). The Theorem is proved
\end{proof}

In Estimations 1-5 we use the notations: $G(+i,a)=\{x\in G,x_{i}>a\},$
$G(-i,a)=\{x\in G,x_{i}<-a\},$ where $x=(x_{1},x_{2},...,x_{d}),a>0.$ It is
not hard to verify that for any subset $G$ of $U_{\varepsilon}(S_{\rho
}^{^{\prime}})\cup U_{2\varepsilon}(A(\rho))$ , that is, for all considered
sets $G$ in these estimations, and for any $x\in G$ the followings hold
\begin{equation}
\rho-1<\mid x\mid<\rho+1,\text{ }G\subset(\cup_{i=1}^{d}(G(+i,\rho d^{-1})\cup
G(-i,\rho d^{-1}))
\end{equation}
Indeed, if $x\in S_{\rho}^{^{\prime}},$ then $F(x)=\rho^{2}$ and by definition
of $F(x)$ ( see Lemma 2) and (23) we have $\mid x\mid=\rho+O(\rho
^{-1-\alpha_{1}}).$ Hence the inequalities in (79) hold for $x\in
U_{\varepsilon}(S_{\rho}^{^{\prime}}).$ If $x\in$ $A(\rho),$ then by
definition of $A(\rho)$ ( see (61), (62)), we have $x\in K_{\rho},$ and hence
$\mid x\mid=\rho+O(\rho^{-1+\alpha_{1}})$. Thus the inequalities in (79) hold
for $x\in U_{2\varepsilon}(A(\rho))$ too. The inclusion in (79) follows from
these inequalities.

If $G$ $\subset S_{\rho},$ then by (43) we have $\frac{\partial F(x)}{\partial
x_{k}}>0$ for $x\in G(+k,\rho^{-\alpha})$. Therefore to calculate the measure
of $G(+k,a)$ for $a\geq\rho^{-\alpha}$ we use the formula
\begin{equation}
\mu(G(+k,a))=\int\limits_{\Pr_{k}(G(+k,a))}(\frac{\partial F}{\partial x_{k}%
})^{-1}\mid grad(F)\mid dx_{1}...dx_{k-1}dx_{k+1}...dx_{d},
\end{equation}
where $\Pr_{k}(G)\equiv\{(x_{1},x_{2},...,x_{k-1},x_{k+1},x_{k+2}%
,...,x_{d}):x\in G\}$ is the projection of $G$ on the hyperplane $x_{k}=0.$
Instead of $\Pr_{k}(G)$ we write $\Pr(G)$ if $k$ is unambiguous. If $D$ is
$m-$dimensional subset of $\mathbb{R}^{m},$ then to estimate $\mu(D),$ we use
the formula
\begin{equation}
\mu(D)=\int\limits_{\Pr_{k}(D)}\mu(D(x_{1},...x_{k-1},x_{k+1},...,x_{m}%
))dx_{1}...dx_{k-1}dx_{k+1}...dx_{m},
\end{equation}
where $D(x_{1},...x_{k-1},x_{k+1},...,x_{m})=\{x_{k}:(x_{1},x_{2}%
,...,x_{m})\in D\}.$

ESTIMATION 1. Here we prove (76) by using (80). During this estimation the set
$S_{\rho}$ is redenoted by $G.$ If $x\in G,$ then $x\notin V_{b}(\rho
^{\alpha_{1}})$ for all $b\in\Gamma(p\rho^{\alpha}).$ Since the rotation does
not change the measure, we choose the coordinate axis so that the direction of
a fixed $b\in\Gamma(p\rho^{\alpha})$ coinsides with the direction of
$(1,0,0,...,0),$ that is, $b=(b_{1},0,0,...,0),b_{1}>0.$ Then the relations

$x\notin V_{b}(\rho^{\alpha_{1}}),$ $\mid b\mid<p\rho^{\alpha},\alpha
_{1}=3\alpha$ imply that $\mid x_{1}\mid>a,$ where

$a=(\rho^{\alpha_{1}}-b_{1}^{2})(2b_{1})^{-1}>\rho^{\alpha}.$ Therefore
$G=G(+1,a)\cup G(-1,a).$ Now we estimate $\mu(G(+1,a))$ by using (80) for
$k=1$ and \ the relations%
\begin{align}
\frac{\partial F}{\partial x_{1}}  &  >\rho^{\alpha},(\frac{\partial
F}{\partial x_{1}})^{-1}\mid grad(F)\mid=\frac{\mid x\mid}{x_{1}}%
+O(\rho^{-2\alpha}),\\
\Pr(G(+1,a))  &  \supset\Pr(A(+1,2a)),
\end{align}
where $x\in G(+1,a),a>\rho^{\alpha},$ $A=B(\rho)\cap U(3\rho^{\alpha_{1}},p).$
Here (82) follows from (43). Now we prove (83). If $(x_{2},...,x_{d})\in
\Pr_{1}(A(+1,2a)),$ then by definition of $A(+1,2a)$ there exists $x_{1}$ such
that
\begin{equation}
x_{1}>2a>2\rho^{\alpha},\text{ }x_{1}^{2}+x_{2}^{2}+...+x_{d}^{2}=\rho
^{2},\mid\sum_{i\geq1}(2x_{i}b_{i}-b_{i}^{2})\mid\geq3\rho^{\alpha_{1}}%
\end{equation}
for all $(b_{1},b_{2},...,b_{d})\in\Gamma(p\rho^{\alpha}).$ Therefore for
$h=\rho^{-\alpha}$ we have

$(x_{1}+h)^{2}+x_{2}^{2}+...+x_{q}^{2}>\rho^{2}+\rho^{-\alpha},(x_{1}%
-h)^{2}+x_{2}^{2}+...+x_{q}^{2}<\rho^{2}-\rho^{-\alpha}.$ This and (23) give
$F(x_{1}+h,x_{2},...,x_{d})>\rho^{2}$, $F(x_{1}-h,x_{2},...,x_{d})<\rho^{2}$.
Since $F$ is a continuous function there is $y_{1}\in(x_{1}-h,x_{1}+h)$ such
that (see (84))
\begin{equation}
y_{1}>a,F(y_{1},x_{2},...,x_{d})=\rho^{2},\mid2y_{1}b_{1}-b_{1}^{2}%
+\sum_{i\geq2}(2x_{i}b_{i}-b_{i}^{2})\mid>\rho^{\alpha_{1}},
\end{equation}
because the expression under the absolute value in (85) differ from the
expression under the absolute value in (84) by $2(y_{1}-x_{1})b_{1},$ where
$\mid y_{1}-x_{1}\mid<h=\rho^{\alpha},$ $b_{1}<p\rho^{\alpha},$ $\mid
2(y_{1}-x_{1})b_{1}\mid<2p\rho^{2\alpha}<$ $\rho^{\alpha_{1}}.$ The relations
in (85) means that $(x_{2},...,x_{d})\in\Pr G(+1,a).$ Hence (83) is proved.
Now (80), (82), and the obvious relation $\mu(\Pr G(+1,a))=O(\rho^{d-1})$ (
see the inequalities in (79)) imply that%
\[
\mu(G(+1,a))=\int\limits_{\Pr(G(+1,a))}\frac{\mid x\mid}{x_{1}}dx_{2}%
dx_{3}...dx_{d}+O(\rho^{-\alpha})\mu(B(\rho))\geq
\]%
\[
\int\limits_{\Pr(A(+1,2a))}\frac{\mid x\mid}{x_{1}}dx_{2}dx_{3}...dx_{d}%
-c_{14}\rho^{-\alpha}\mu(B(\rho))=
\]%
\[
\mu(A(+1,2a))-c_{14}\rho^{-\alpha}\mu(B(\rho)).
\]
Similarly, $\mu(G(-1,a))\geq\mu(A(-1,2a))-c_{14}\rho^{-\alpha}\mu(B(\rho)).$
Therefore using the relations $G=G(-1,a)\cup G(+1,a)),$ $A=A(-1,2a)\cup
A(+1,2a),$

$\mu(A))=(1+O(\rho^{-\alpha}))\mu(B(\rho))$ (see (41) ) we obtain (76).

ESTIMATION 2 Here we prove (77). For this we estimate the measure of the set
$S_{\rho}\cap P_{b}$ by using (80). During this estimation the set $S_{\rho
}\cap P_{b}$ is redenoted by $G$. We choose the coordinate axis so that the
direction of $b$ coincides with the direction of $(1,0,0,...,0),$ i.e.,
$b=(b_{1},0,0,...,0)$ and $b_{1}>0$. It follows from the definitions of
$S_{\rho},P_{b}$ and $F(x)$ ( see the beginning of this section, (60), and
Lemma 2(a)) that if $(x_{1},x_{2},...,x_{d})\in G$ then
\begin{align}
x_{1}^{2}+x_{2}^{2}+...+x_{d}^{2}+F_{k_{1}-1}(x)  &  =\rho^{2},\\
(x_{1}+b_{1})^{2}+x_{2}^{2}+x_{3}^{2}+...+x_{d}^{2}+F_{k_{1}-1}(x+b)  &
=\rho^{2}+h,
\end{align}
where $h\in(-3\varepsilon_{1},3\varepsilon_{1}).$ Subtracting (86) from (87)
and using (23), we get
\begin{equation}
(2x_{1}+b_{1})b_{1}=O(\rho^{-\alpha_{1}}).
\end{equation}
This and the inequalities in (79) imply
\begin{equation}
\mid b_{1}\mid<2\rho+3,\text{ }x_{1}=\frac{b_{1}}{2}+O(\rho^{-\alpha_{1}}%
b_{1}^{-1}),\mid x_{1}^{2}-(\frac{b_{1}}{2})^{2}\mid=O(\rho^{-\alpha_{1}}).
\end{equation}
Consider two cases. Case 1: $b\in\Gamma_{1},$ where $\Gamma_{1}=\{b\in
\Gamma:\mid\rho^{2}-\mid\frac{b}{2}\mid^{2}\mid<3d\rho^{-2\alpha}\}.$ In this
case using the last equality in (89), (86), (23), and taking into account that
$b=(b_{1},0,0,...,0),$ $\alpha_{1}=3\alpha,$ we obtain
\begin{equation}
x_{1}^{2}=\rho^{2}+O(\rho^{-2\alpha}),\mid x_{1}\mid=\rho+O(\rho^{-2\alpha
-1}),x_{2}^{2}+x_{3}^{2}+...+x_{d}^{2}=O(\rho^{-2\alpha}).
\end{equation}
Therefore $G\subset G(+1,a)\cup G(-1,a),$ where $a=\rho-\rho^{-1}$. Using
(80), the obvious relation $\mu(\Pr_{1}(G(\pm1,a))=O(\rho^{-(d-1)\alpha})$
(see (90)) and taking into account that the expression under the integral in
(80) for $k=1$ is equal to $1+O(\rho^{-\alpha})$ (see (82) and (90)), we get
$\mu(G(\pm1,a))=O(\rho^{-(d-1)\alpha}).$ Thus $\mu(G)=O(\rho^{-(d-1)\alpha}).$
Since $\mid\Gamma_{1}\mid=O(\rho^{d-1}),$ we have
\begin{equation}
\text{ }\mu(\cup_{b\in\Gamma_{1}}(S_{\rho}\cap P_{b})=O(\rho^{-(d-1)\alpha
+d-1})=O(\rho^{-\alpha})\mu(B(\rho)).
\end{equation}
Case 2: $\mid\rho^{2}-\mid\frac{b}{2}\mid^{2}\mid\geq3d\rho^{-2\alpha}.$
Repeating the proof of (90), we get%
\begin{equation}
\mid x_{1}^{2}-\rho^{2}\mid>2d\rho^{-2\alpha},\text{ }\sum_{k=2}^{d}x_{k}%
^{2}>d\rho^{-2\alpha},\text{ }\max_{k\geq2}\mid x_{k}\mid>\rho^{-\alpha}.
\end{equation}
Therefore $G\subset\cup_{k\geq2}(G(+k,\rho^{-\alpha})\cup G(-k,\rho^{-\alpha
})).$ Now we estimate $\mu(G(+d,\rho^{-\alpha}))$ by using (80). Redenote by
$D$ the set $\Pr_{d}G(+d,\rho^{-\alpha}).$ If $x\in G(+d,\rho^{-\alpha}),$
then according to (86) and (43) the under integral expression in (80) for
$k=d$ is $O(\rho^{1+\alpha}).$ Therefore the first equality in
\begin{equation}
\mu(D)=O(\varepsilon_{1}\mid b\mid^{-1}\rho^{d-2}),\text{ }\mu(G(+d,\rho
^{-\alpha}))=O(\rho^{d-1+\alpha}\varepsilon_{1}\mid b\mid^{-1})
\end{equation}
implies the second equality in (93). To prove the first equality in (93) we
use (81) for $m=d-1$ and $k=1$ and prove the relations $\mu(\Pr_{1}%
D)=O(\rho^{d-2}),$
\begin{equation}
\text{ }\mu(D(x_{2},x_{3},...,x_{d-1}))<6\varepsilon_{1}\mid b\mid^{-1}%
\end{equation}
for $(x_{2},x_{3},...,x_{d-1})\in\Pr_{1}D.$ First relation follows from the
inequalities in (79)). So we need to prove (94). If $x_{1}\in D(x_{2}%
,x_{3},...,x_{d-1})$ then (86) and (87) holds. Subtructing (86) from (87), we
get
\begin{equation}
2x_{1}b_{1}+(b_{1})^{2}+F_{k_{1}-1}(x-b)-F_{k_{1}-1}(x)=h,
\end{equation}
where $x_{2},x_{3},...,x_{d-1}$ are fixed . Hence we have two equations (86)
and (95) with respect two unknown $x_{1}$ and $x_{d}$. Using (43), the
implicit function theorem, and the inequalities $\mid x_{d}\mid>\rho^{-\alpha
},$ $\alpha_{1}>2\alpha$ from (86), we obtain
\begin{equation}
x_{d}=f(x_{1}),\text{ }\frac{df}{dx_{1}}=-\frac{2x_{1}+O(\rho^{-2\alpha
_{1}+\alpha})}{2x_{d}+O(\rho^{-2\alpha_{1}+\alpha})}=-\frac{x_{1}}{x_{d}%
}+O(\rho^{-\alpha_{1}}).
\end{equation}
Substituting this in (95), we get
\begin{equation}
2x_{1}b_{1}+b_{1}^{2}+F_{k_{1}-1}(x_{1}+b_{1},x_{2},...,x_{d-1},f(x_{1}%
))-F_{k_{1}-1}(x_{1},...,x_{d-1},f)=h.
\end{equation}
Using (43), (96), the first equality in (89), and $x_{d}>\rho^{-\alpha}$ we
see that the absolute value of the derivative (w.r.t. $x_{1}$) of the
left-hand side of (97) satisfies

$\mid2b_{1}+O(\rho^{-2\alpha_{1}+\alpha})(1+\mid\frac{df}{dx_{1}}\mid)\mid=$

$\mid2b_{1}+O(\rho^{-2\alpha_{1}+\alpha})(1+\mid\frac{x_{1}}{x_{d}}%
\mid)+O(\rho^{-3\alpha_{1}+\alpha})\mid>b_{1}.$ Therefore from (97) by
implicit function theorem, we get $\mid\frac{dx_{1}}{dh}\mid<\frac{1}{\mid
b\mid}.$ This inequality and relation $h\in(-3\varepsilon_{1},3\varepsilon
_{1})$ imply (94). Thus (93) is proved. In the same way we get the same
estimation for $G(+k,\rho^{-\alpha})$ and $G(-k,\rho^{-\alpha})$ for $k\geq2$. Hence

$\mu(S_{\rho}\cap P_{b})=O(\rho^{d-1+\alpha}\varepsilon_{1}\mid b\mid^{-1}),$
for $b\notin\Gamma_{1}.$ Since $\mid b\mid<2\rho+3$ ( see (89)) and
$\varepsilon_{1}=\rho^{-d-2\alpha}$, taking into account that the number of
the vectors of $\Gamma$ satisfying $\mid b\mid<2\rho+3$ is $O(\rho^{d}),$ we obtain

$\mu(\cup_{b\notin\Gamma_{1}}(S_{\rho}\cap P_{b}))=O(\rho^{2d-1+\alpha
}\varepsilon_{1})=O(\rho^{-\alpha})\mu(B(\rho)).$ This, (91) and (76) give the
proof of (77).

ESTIMATION 3. Here we prove (78). Denote $U_{2\varepsilon}(A_{k,j}(\gamma
_{1,}\gamma_{2},...,\gamma_{k}))$ by $G,$ where $\gamma_{1,}\gamma
_{2},...,\gamma_{k}\in\Gamma(p\rho^{\alpha}),k\leq d-1,$ and $A_{k,j}$ is
defined in (61). We turn the coordinate axis so that

$Span\{\gamma_{1,}\gamma_{2},...,\gamma_{k}\}=\{x=(x_{1},x_{2},...,x_{k}%
,0,0,...,0):x_{1},x_{2},...,x_{k}\in\mathbb{R}\}$. Then by (31), we have
$x_{n}=O(\rho^{\alpha_{k}+(k-1)\alpha})$ for $n\leq k,$ $x\in G$. This, (79), and

$\alpha_{k}+(k-1)\alpha<1$ ( see the sixth inequality in (14)) give

$G\subset(\cup_{i>k}(G(+i,\rho d^{-1})\cup G(-i,\rho d^{-1})),$

$\mu(\Pr_{i}(G(+i,\rho d^{-1})))=O(\rho^{k(\alpha_{k}+(k-1)\alpha)+(d-1-k)})$
for $i>k.$ Now using this and (81) for $m=d,$ we prove that
\begin{equation}
\mu(G(+i,\rho d^{-1}))=O(\varepsilon\rho^{k(\alpha_{k}+(k-1)\alpha
)+(d-1-k)}),\forall i>k.
\end{equation}
For this we redenote by $D$ the set $G(+i,\rho d^{-1})$ and prove that
\begin{equation}
\mu((D(x_{1},x_{2},...x_{i-1},x_{i+1},...x_{d}))\leq(42d^{2}+4)\varepsilon
\end{equation}
for $(x_{1},x_{2},...x_{i-1},x_{i+1},...x_{d})\in\Pr_{i}(D)$ and $i>k.$ To
prove (99) it is sufficient to show that if both $x=(x_{1},x_{2}%
,...,x_{i},...x_{d})$ and $x^{^{\prime}}=(x_{1},x_{2},...,x_{i}^{^{\prime}%
},...,x_{d})$ are in $D,$ then $\mid x_{i}-x_{i}^{^{\prime}}\mid\leq
(42d^{2}+4)\varepsilon.$ Assume the converse. Then $\mid x_{i}-x_{i}%
^{^{\prime}}\mid>(42d^{2}+4)\varepsilon$. Without loss of generality it can be
assumed that $x_{i}^{^{\prime}}>x_{i}.$ So $x_{i}^{^{\prime}}>x_{i}>\rho
d^{-1}$ ( see definition of $D$). Since $x$ and $x^{^{\prime}}$ lie in the
$2\varepsilon$ neighborhood of $A_{k,j},$ there exist points $a$ and
$a^{^{\prime}}$ in $A_{k,j}$ such that $\mid x-a\mid<2\varepsilon$ and $\mid
x^{^{\prime}}-a^{^{\prime}}\mid<2\varepsilon.$ It follows from the definitions
of the points $x,$ $x^{^{\prime}},a,$ $a^{^{\prime}}$ that the following
inequalities hold:%
\begin{align}
\rho d^{-1}-2\varepsilon &  <a_{i}<a_{i}^{^{\prime}},\text{ }a_{i}^{^{\prime}%
}-a_{i}>42d^{2}\varepsilon,\\
(a_{i}^{^{\prime}})^{2}-(a_{i})^{2}  &  >2(\rho d^{-1}-2\varepsilon
)(a_{i}^{^{\prime}}-a_{i}),\nonumber\\
&  \mid\mid a_{s}\mid-\mid a_{s}^{^{\prime}}\mid\mid<4\varepsilon,\forall
s\neq i.\nonumber
\end{align}
On the other hand for points of $A_{k,j}$ the inequalities in (79) hold, that
is, we have $\mid a_{s}\mid<\rho+1,\mid a_{s}^{^{\prime}}\mid<\rho+1.$
Therefore these inequalities and the inequalities in (100) imply $\mid\mid
a_{s}\mid^{2}-\mid a_{s}^{^{\prime}}\mid^{2}\mid<12\rho\varepsilon$ for $s\neq
i$, and hence

$\sum_{s\neq i}\mid\mid a_{s}\mid^{2}-\mid a_{s}^{^{\prime}}\mid^{2}%
\mid<12d\rho\varepsilon<\frac{2}{7}\rho d^{-1}(a_{i}^{^{\prime}}-a_{i}),$%

\begin{equation}
\mid\mid a\mid^{2}-\mid a^{^{\prime}}\mid^{2}\mid>\frac{3}{2}\rho d^{-1}\mid
a_{i}^{^{\prime}}-a_{i}\mid.
\end{equation}
Now using the inequality (45), the obvious relation $\frac{1}{2}\alpha_{d}<1$
( see the end of the introduction), the notations $r_{j}(x)=\lambda
_{j}(x)-\mid x\mid^{2}$ ( see Remark 2), $\varepsilon_{1}=7\rho\varepsilon$ (
see Lemma 2(a)), and (101), (100), we get

$\mid r_{j}(a)-r_{j}(a^{^{\prime}})\mid<\rho^{\frac{1}{2}\alpha_{d}}\mid
a-a^{^{\prime}}\mid<\frac{1}{2}\rho d^{-1}\mid a_{i}^{^{\prime}}-a_{i}\mid,$

$\mid\lambda_{j}(a)-\lambda_{j}(a^{^{\prime}})\mid\geq\mid\mid a\mid^{2}-\mid
a^{^{\prime}}\mid^{2}\mid-\mid r_{j}(a)-r_{j}(a^{^{\prime}})\mid>$

$\rho d^{-1}\mid a_{i}^{^{\prime}}-a_{i}\mid>42d\rho\varepsilon>6\varepsilon
_{1}.$

The obtained inequality $\mid\lambda_{j}(a)-\lambda_{j}(a^{^{\prime}}%
)\mid>6\varepsilon_{1}$ controdicts with inclutions $a$ $\in A_{k,j},$
$a^{^{\prime}}\in A_{k,j},$ since by definition of $A_{k,j}$ ( see (61)) both
$\lambda_{j}(a)$ and $\lambda_{j}(a^{^{\prime}})$ lie in $(\rho^{2}%
-3\varepsilon_{1},\rho^{2}+3\varepsilon_{1}).$ Thus (99), hense (98) is
proved. In the same way we get the same formula for $G(-i,\frac{\rho}{d}).$ So

$\mu(U_{2\varepsilon}(A_{k,j}(\gamma_{1,}\gamma_{2},...,\gamma_{k}%
)))=O(\varepsilon\rho^{k(\alpha_{k}+(k-1)\alpha)+d-1-k}).$ Now taking into
account that $U_{2\varepsilon}(A(\rho))$ is union of $U_{2\varepsilon}%
(A_{k,j}(\gamma_{1,}\gamma_{2},...,\gamma_{k})$ for $k=1,2,..,d-1;$
$j=1,2,...,b_{k}(\gamma_{1},\gamma_{2},...,\gamma_{k}),$ and $\gamma
_{1},\gamma_{2},...,\gamma_{k}\in\Gamma(p\rho^{\alpha})$ \ ( see (61)) and
using that $b_{k}=O(\rho^{d\alpha+\frac{k}{2}\alpha_{k+1}})$ ( see (39)) and
the number of the vectors $(\gamma_{1},\gamma_{2},...,\gamma_{k})$ for
$\gamma_{1},\gamma_{2},...,\gamma_{k}\in\Gamma(p\rho^{\alpha})$ is
$O(\rho^{dk\alpha}),$ we obtain
\[
\mu(U_{2\varepsilon}(A(\rho)))=O(\varepsilon\rho^{d\alpha+\frac{k}{2}%
\alpha_{k+1}+dk\alpha+k(\alpha_{k}+(k-1)\alpha)+d-1-k}).
\]
Therefore to prove (78), it remains to show that

$d\alpha+\frac{k}{2}\alpha_{k+1}+dk\alpha+k(\alpha_{k}+(k-1)\alpha)+d-1-k\leq
d-1-\alpha$ or%
\begin{equation}
(d+1)\alpha+\frac{k}{2}\alpha_{k+1}+dk\alpha+k(\alpha_{k}+(k-1)\alpha)\leq k
\end{equation}
for $1\leq k\leq d-1$. Dividing both side of (102) by $k\alpha$ and using
$\alpha_{k}=3^{k}\alpha,$ $\alpha=\frac{1}{q},$ $q=3^{d}+d+2$ ( see the end of
the introduction) we see that (102) is equivalent to $\frac{d+1}{k}%
+\frac{3^{k+1}}{2}+3^{k}+k-1\leq3^{d}+2.$ The left-hand side of this
inequality gets its maximum at $k=d-1.$ Therefore we need to show that
$\frac{d+1}{d-1}+\frac{5}{6}3^{d}+d\leq3^{d}+4,$ which follows from the
inequalities $\frac{d+1}{d-1}\leq3,d<\frac{1}{6}3^{d}+1$ for $d\geq2.$

ESTIMATION 4. Here we prove (67). During this estimation we denote by $G$ the
set $S_{\rho}^{^{\prime}}\cap U_{\varepsilon}(Tr(A(\rho))$. Since $V_{\rho
}=S_{\rho}^{^{\prime}}\backslash G$ and (77) holds, it is enough to prove that
$\mu(G)=O(\rho^{-\alpha})\mu(B(\rho)).$ For this we use (79) and prove
$\mu(G(+i,\rho d^{-1}))=O(\rho^{-\alpha})\mu(B(\rho))$ for $i=1,2,...,d$ by
using (80) ( the same estimation for $G(-i,\rho d^{-1})$ can be proved in the
same way). By (43), if $x\in G(+i,\rho d^{-1}),$ then the under integral
expression in (80) for $k=i$ and $a=\rho d^{-1}$ is less than $d+1.$ Therefore
it is sufficient to prove
\begin{equation}
\mu(\Pr(G(+i,\rho d^{-1}))=O(\rho^{-\alpha})\mu(B(\rho))
\end{equation}
Clearly, if $(x_{1},x_{2},...x_{i-1},x_{i+1},...x_{d})\in\Pr_{i}(G(+i,\rho
d^{-1})),$ then

$\mu(U_{\varepsilon}(G)(x_{1},x_{2},...x_{i-1},x_{i+1},...x_{d}))\geq
2\varepsilon$ and by (81), it follows that
\begin{equation}
\mu(U_{\varepsilon}(G))\geq2\varepsilon\mu(\Pr(G(+i,\rho d^{-1})).
\end{equation}
Hence to prove (103) we need to estimate $\mu(U_{\varepsilon}(G)).$ For this
we prove that
\begin{equation}
U_{\varepsilon}(G)\subset U_{\varepsilon}(S_{\rho}^{^{\prime}}),U_{\varepsilon
}(G)\subset U_{2\varepsilon}(Tr(A(\rho))),U_{\varepsilon}(G)\subset
Tr(U_{2\varepsilon}(A(\rho))).
\end{equation}
The first and second inclusions follow from $G\subset S_{\rho}^{^{\prime}}$
and $G\subset U_{\varepsilon}(Tr(A(\rho)))$ respectively (see definition of
$G$ ). Now we prove the third inclusion in (105). If $x\in U_{\varepsilon
}(G),$ then by the second inclusion of (105) there exists $b$ such that $b\in
Tr(A(\rho)),$ $\mid x-b\mid<2\varepsilon.$ Then by the definition of
$Tr(A(\rho))$ there exist $\gamma\in\Gamma$ and $c\in A(\rho)$ such that
$b=\gamma+c$. Therefore $\mid x-\gamma-c\mid=\mid x-b\mid<2\varepsilon,$

$x-\gamma\in U_{2\varepsilon}(c)\subset U_{2\varepsilon}(A(\rho)).$ This
together with $x\in U_{\varepsilon}(G)\subset U_{\varepsilon}(S_{\rho
}^{^{\prime}})$ (see the first inclusion of (105)) give $x\in
Tr(U_{2\varepsilon}(A(\rho)))$ ( see the definition of $Tr(E)$ in the
beginning of this section), i.e., the third inclusion in (105) is proved. The
third inclusion, Lemma 2(c), and (78) imply that

$\mu(U_{\varepsilon}(G))=O(\rho^{-\alpha})\mu(B(\rho))\varepsilon.$ This and
(104) imply the proof of (103)$\diamondsuit$

ESTIMATION 5 Here we prove (68). Divide the set $V_{\rho}\equiv V$ into
pairwise disjoint subsets $V^{^{\prime}}(\pm1,\rho d^{-1})\equiv V(\pm1,\rho
d^{-1}),$

$V^{^{\prime}}(\pm i,\rho d^{-1})\equiv V(\pm i,\rho d^{-1})\backslash
(\cup_{j=1}^{i-1}(V(\pm j,\rho d^{-1}))),$ for $i=2,3,...,d.$ Take any point
$a\in V^{^{\prime}}(+i,\rho d^{-1})\subset S_{\rho}$ and consider the function
$F(x)$ ( see Lemma 2(a)) on the interval $[a-\varepsilon e_{i},a+\varepsilon
e_{i}],$ where $e_{1}=(1,0,0,...,0)$, $e_{2}=(0,1,0,...,0),...$. By the
definition of \ $S_{\rho}$ we have $F(a)=\rho^{2}$. It follows from (43) and
the definition of $V^{^{\prime}}(+i,\rho d^{-1})$ that $\frac{\partial
F(x)}{\partial x_{i}}>\rho d^{-1}$ for $x\in\lbrack a-\varepsilon
e_{i},a+\varepsilon e_{i}].$ Therefore
\begin{equation}
F(a-\varepsilon e_{i})<\rho^{2}-c_{15}\varepsilon_{1},\text{ }F(a+\varepsilon
e_{i})>\rho^{2}+c_{15}\varepsilon_{1}.
\end{equation}
Since $[a-\varepsilon e_{i},a+\varepsilon e_{i}]\in U_{\varepsilon}(a)\subset
U_{\varepsilon}(V_{\rho})\subset U_{\varepsilon}(S_{\rho}^{^{\prime}%
})\backslash Tr(A(\rho))$ ( see Theorem 5$(b)$), it follows from Theorem
5$(a)$ that there exists index $N$ such that $\Lambda(y)=\Lambda_{N}(y)$ for
$y\in U_{\varepsilon}(a)$ and $\Lambda(y)$ satisfies (46)( see Remark 3).
Hence (106) implies that
\begin{equation}
\Lambda(a-\varepsilon e_{i})<\rho^{2},\Lambda(a+\varepsilon e_{i})>\rho^{2}.
\end{equation}
Moreover it follows from (66) that the derivative of $\Lambda(y)$ with respect
to $i$-th coordinate is positive for $y\in\lbrack a-\varepsilon e_{i}%
,a+\varepsilon e_{i}].$ So $\Lambda(y)$ is a continuous and increasing
function in $[a-\varepsilon e_{i},a+\varepsilon e_{i}].$ Therefore (107)
implies that there exists a unique point $y(a,i)\in\lbrack a-\varepsilon
e_{i},a+\varepsilon e_{i}]$ such that $\Lambda(y(a,i))=\rho^{2}.$ Define
$I_{\rho}^{^{\prime}}(+i)$ by $I_{\rho}^{^{\prime}}(+i)=\{y(a,i):a\in
V^{^{\prime}}(+i,\rho d^{-1})\}).$ In the same way we define $I_{\rho
}^{^{\prime}}(-i)=\{y(a,i):a\in V^{^{\prime}}(-i,\rho d^{-1})\}$ and put
$I_{\rho}^{^{\prime}}=\cup_{i=1}^{d}(I_{\rho}^{^{\prime}}(+i)\cup I_{\rho
}^{^{\prime}}(-i)).$ To estimate the measure of $I_{\rho}^{^{\prime}}$ we
compare the measure of $V^{^{\prime}}(\pm i,\rho d^{-1})$ with the measure of
$I_{\rho}^{^{\prime}}(\pm i)$ by using the formula (80) and the obvious
relations
\begin{equation}
\Pr(V^{^{\prime}}(\pm i,\rho d^{-1}))=\Pr(I_{\rho}^{^{\prime}}(\pm i)),\text{
}\mu(\Pr(I_{\rho}^{^{\prime}}(\pm i)))=O(\rho^{d-1}),
\end{equation}%
\begin{equation}
(\frac{\partial F}{\partial x_{i}})^{-1}\mid grad(F)\mid-(\frac{\partial
\Lambda}{\partial x_{i}})^{-1}\mid grad(\Lambda)\mid=O(\rho^{-2\alpha_{1}}).
\end{equation}
Here the first equality in (108) follows from the definition of $I_{\rho
}^{^{\prime}}(\pm i)$. The second equality in (108) follows from the
inequalities in (79), since $I_{\rho}^{^{\prime}}\subset U_{\varepsilon
}(S_{\rho}^{^{\prime}})$. Formulas (43), (66) imply (109). Clearly, using
(108), (109), and (80) we get $\mu(V^{^{\prime}}(\pm i,\rho d^{-1}%
))-\mu(I_{\rho}^{^{\prime}}(\pm i))=O(\rho^{d-1-2\alpha_{1}}).$ On the other
hand if

$y\equiv(y_{1},y_{2},...,y_{d})\in I_{\rho}^{^{\prime}}(+i)\cap I_{\rho
}^{^{\prime}}(+j)$ for $i<j$ then there are $a\in V^{^{\prime}}(+i,\rho
d^{-1})$ and $a^{^{\prime}}\in V^{^{\prime}}(+j,\rho d^{-1})$ such that
$y=y(a,i)=y(a^{^{\prime}},j)$ and $y\in\lbrack a-\varepsilon e_{i}%
,a+\varepsilon e_{i}],$ $y\in\lbrack a^{^{\prime}}-\varepsilon e_{j}%
,a^{^{\prime}}+\varepsilon e_{j}].$ These inclusions imply that $\rho
d^{-1}-\varepsilon\leq y_{i}\leq\rho d^{-1}.$ Therefore $\mu(\Pr_{j}(I_{\rho
}^{^{\prime}}(+i)\cap I_{\rho}^{^{\prime}}(+j)))=O(\varepsilon\rho^{d-2}).$
This equality, (80) and (66) imply that $\mu((I_{\rho}^{^{\prime}}(+i)\cap
I_{\rho}^{^{\prime}}(+j)))=O(\varepsilon\rho^{d-2})$ for all $i$ and $j.$ Similarly

$\mu((I_{\rho}^{^{\prime}}(+i)\cap I_{\rho}^{^{\prime}}(-j)))=O(\varepsilon
\rho^{d-2})$ for all $i$ and $j.$ Thus%
\[
\mu(I_{\rho}^{^{\prime}})=\sum_{i}\mu(I_{\rho}^{^{\prime}}(+i))+\sum_{i}%
\mu(I_{\rho}^{^{\prime}}(-i))+O(\varepsilon\rho^{d-2})=
\]
$\ \ \ \ \ \ \ \ \ \ \ \ \ \ \sum_{i}\mu(V^{^{\prime}}(+i,\rho d^{-1}%
))+\sum_{i}\mu(V^{^{\prime}}(-i,\rho d^{-1}))+O(\rho^{d-1-2\alpha_{1}})=$

$\mu(V_{\rho})+O(\rho^{-2\alpha_{1}})\mu(B(\rho)).$ This and (67) yeild the
inequality (68) for $I_{\rho}^{^{\prime}}$. Now we define $I_{\rho}%
^{^{\prime\prime}}$ as follows. If $\gamma+t\in$ $I_{\rho}^{^{\prime}}$ then
$\Lambda(\gamma+t)=\rho^{2}$, where $\Lambda(\gamma+t)$ is a unique eigenvalue
satisfying (5) ( see Remark 3). Since

$\Lambda(\gamma+t)=\mid\gamma+t\mid^{2}+O(\rho^{-\alpha_{1}})$ ( see (5) and
(23)), for fixed $t$ there exist only a finite number of vectors $\gamma
_{1},\gamma_{2},...,\gamma_{s}\in\Gamma$ satisfying $\Lambda(\gamma
_{k}+t)=\rho^{2}$. Hence $I_{\rho}^{^{\prime}}$ is the union of pairwise
disjoint subsets $I_{\rho,k}^{^{\prime}}$ $\equiv\{\gamma_{k}+t\in I_{\rho
}^{^{\prime}}:\Lambda(\gamma_{k}+t)=\rho^{2}\}$ for

$k=1,2,...s.$ The translation $I_{\rho,k}^{^{\prime\prime}}=I_{\rho
,k}^{^{\prime}}-\gamma_{k}=\{t\in F^{\ast}:\gamma_{k}+t\in I_{\rho
,k}^{^{\prime}}\}$ of $I_{\rho,k}^{^{\prime}}$ is a part of the isoenergetic
surfaces $I_{\rho}$ of $L(q(x)).$ Put $I_{\rho}^{^{\prime\prime}}=\cup
_{k=1}^{s}I_{\rho,k}^{^{\prime\prime}}.$ If $t\in I_{\rho,k}^{^{\prime\prime}%
}\cap I_{\rho,m}^{^{\prime\prime}}$ for $k\neq m,$ then $\gamma_{k}+t\in
I_{\rho}^{^{\prime}}\subset U_{\varepsilon}(S_{\rho}^{^{\prime}})$ and
$\gamma_{m}+t\in U_{\varepsilon}(S_{\rho}^{^{\prime}}),$ which contradict
Lemma 2(b). So $I_{\rho}^{^{\prime\prime}}$ is union of the pairwise disjoint
subsets $I_{\rho,k}^{^{\prime\prime}}$ for $k=1,2,...s.$ Thus

$\mu(I_{\rho}^{^{\prime\prime}})=\sum_{k}\mu(I_{\rho,k}^{^{\prime\prime}%
})=\sum_{k}\mu(I_{\rho,k}^{^{\prime}})=\mu(I_{\rho}^{^{\prime}})>(1-c_{10}%
\rho^{-\alpha}))\mu(B(\rho))\diamondsuit$

\end{document}